\documentclass[twocolumn,aps,pr,superscriptaddress,preprintnumbers,nofootinbib,10pt]{revtex4-2}
\usepackage{amsmath,amssymb}
\usepackage[dvipdf,dvips]{graphicx}
\usepackage{color}
\usepackage{hyperref}
\usepackage{url}
\usepackage{slashed}
\usepackage{subfigure}
\usepackage{amsmath}
\usepackage{amsfonts}
\usepackage{float} 
\usepackage{amssymb}
\usepackage{epsfig}
\usepackage{graphics}
\usepackage{euscript}
\usepackage{slashed}
\usepackage{epstopdf}
\usepackage[utf8]{inputenc}
\allowdisplaybreaks
\usepackage{pifont}
\usepackage{dsfont}
\usepackage{MnSymbol}
\usepackage{verbatim}
\usepackage{graphicx}
\usepackage{latexsym}
\usepackage{courier}
\usepackage{multirow}

\usepackage{tikz-feynman}

\def \pt{\partial}

\def \STr{\textmd{STr}}
\def \and{\textmd{and}}

\def \chSB{$\chi$SB}

\graphicspath{{./}}
\newbox{\ORCIDicon}
\sbox{\ORCIDicon}{\large\includegraphics[width=0.8em]{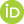}}

\begin{document}

	\title{Quadratic gravity in analogy to quantum chromodynamics:\\
		Light fermions in its landscape}
	
	\author{Gustavo P. de Brito \href{https://orcid.org/0000-0003-2240-528X}{\usebox{\ORCIDicon}}}
	\email{gustavo@cp3.sdu.dk}
	\affiliation{CP3-Origins, University of Southern Denmark, Campusvej 55, DK-5230 Odense M, Denmark}

	\begin{abstract}
		We investigate a non-perturbative approach to quantum gravity built in terms of analogies between quadratic gravity and quantum chromodynamics.
		This approach is based on a conjectured phase transition between quadratic gravity in the trans-Planckian regime and an effective field theory, with general relativity as the leading-order term, below the Planck scale.
		We point out that not all aspects of the analogy between quadratic gravity and quantum chromodynamics are desired. A possible mechanism of chiral symmetry breaking driven by quantum gravity fluctuations could make this setup incompatible with our observed Universe.
		Here, we put forward a first investigation of chiral symmetry breaking in the context of quadratic gravity. We find indication that gravity, despite being an attractive force, does not trigger chiral symmetry breaking in a non-perturbative regime. 
		This result is based on a (functional) renormalization group analysis of four-fermion interactions coupled to quadratic gravity.
		We also comment on the particularities associated with the single-flavor case.
		In summary, the analogy between quadratic gravity and chromodynamics passes this first phenomenological viability test.
	\end{abstract}

	\maketitle
	
	%%%%%%%%%%%%%%%%%%%%%%%%%%%%%%%%%%%%%%%%%%%%%%%%%%%%%%%%%%%%%%%%%%%%%%%%%%%%%%%%%%%%%%%%%%%	
	\section{Introduction}
	
	Since the seminal work by K. Stelle \cite{Stelle:1976gc}, it is known that quadratic gravity, \textit{i.e.}, the Einstein-Hilbert action augmented by curvature squared terms, is renormalizable to all orders in perturbation theory. The underlying reason is that the addition of curvature squared terms enhances the graviton propagator from $p^{-2}$ scaling to $p^{-4}$ scaling, therefore reducing the degree of divergence of Feynman diagrams involving graviton lines. A few years after the proof of renormalizability of quadratic gravity, it was also noted that couplings associated with the curvature squared terms admit asymptotically free trajectories \cite{Julve:1978xn,Tomboulis:1980bs,Fradkin:1981iu,Fradkin:1981hx,Avramidi:1985ki,Avramidi:1986mj,deBerredo-Peixoto:2003jda,deBerredo-Peixoto:2004cjk,Codello:2006in,Narain:2011gs,Narain:2012nf,Ohta:2013uca,Salvio:2018crh}.
	
	At first glance, the properties of all-order renormalizability and asymptotic freedom seem to make quadratic gravity a perfect contender in the quest for a UV-complete theory of quantum gravity. Unfortunately, quadratic gravity has a longstanding issue related to unitarity \cite{Stelle:1976gc,Tomboulis:1983sw,Antoniadis:1986tu,Johnston:1987ue}. Schematically, the tree-level graviton propagator of quadratic gravity behaves as
	\begin{equation}
		\mathcal{D}_{hh}(q^2) \sim -\frac{i}{q^2(q^2-m^2)}\sim \frac{1}{m^2} \bigg( \frac{i}{q^2} - \frac{i}{q^2 - m^2} \bigg) \,,
	\end{equation}
	where $m^2$ is a massive parameter (proportional to the Planck mass). According to the usual ``rules'' of perturbative quantum field theory, the negative sign in the last term of the propagator leads to states with negative norm, \textit{i.e.}, a \textit{ghost}. Since this ghost arises as a consequence of the higher-derivative terms in the action of quadratic gravity, we name it \textit{higher-derivative ghost}.
	In contrast to Faddeev-Popov ghosts in gauge theories, we cannot remove the higher-derivative ghost from the physical spectrum by means of standard quantization methods. Therefore, in its usual incarnation, the higher-derivative ghost spoils the unitarity of quadratic gravity.  
	
	The unitarity problem in quadratic gravity has a classical counterpart, namely, Ostrogradsky instabilities \cite{Ostrogradsk:1850,Woodard:2015zca}. This type of instability is common in theories containing higher-derivative terms and it is usually used to disregard higher-derivative theories. Nevertheless, it has been recently shown that, even at the non-linear level, it is possible to construct stable classical solutions in quadratic gravity \cite{Held:2023aap}.
	Classical aspects of quadratic gravity were also explored in \cite{Stelle:1977ry,Holdom:2002xy,Lu:2015psa,Lu:2015cqa,Lu:2017kzi,Stelle:2017bdu,Bonanno:2019rsq,Bonanno:2021zoy,Held:2022abx}.
	
	Given the problem with unitarity in quadratic gravity and considering the appearance of many alternative ideas to quantize gravity \cite{Oriti:2009zz,Handbook:QG}, quadratic gravity was, for a long time, left aside in the quest for a UV-complete theory of quantum gravity.
	
	In recent years, there was a number of promising ideas to rehabilitate quadratic gravity as a serious contender for a UV-complete theory of quantum gravity \cite{Salvio:2014soa,Salvio:2015gsi,Salvio:2017qkx,Salvio:2018crh,Mannheim:2006rd,Bender:2007wu,Mannheim:2011ds,Mannheim:2018ljq,Mannheim:2020ryw,Mannheim:2021oat,Donoghue:2018izj,Anselmi:2018bra,Anselmi:2018ibi,Anselmi:2018kgz,Anselmi:2018tmf,Anselmi:2019nie,Anselmi:2019xac,Piva:2023bcf,Donoghue:2019ecz,Donoghue:2019fcb,Donoghue:2021eto,Donoghue:2021meq,Donoghue:2021cza}. 
	Among the different proposals, there are claims that quadratic gravity can be formulated as a unitary theory by bringing back ideas from the Lee-Wick program for higher-derivative theories \cite{Lee:1969fy,Anselmi:2017lia,Anselmi:2017yux,Donoghue:2018lmc}.
	For example, Anselmi and Piva introduced a new prescription to define the integration contour around propagator poles in the complex plane, which allows us to remove the higher-derivative ghost from the physical spectrum
	\cite{Anselmi:2018bra,Anselmi:2018ibi,Anselmi:2018kgz,Anselmi:2018tmf,Anselmi:2019nie,Anselmi:2019xac,Piva:2023bcf}. In this approach, the higher-derivative ghost is interpreted as a fake-particle, or a \textit{fakeon}.
	As a second example, Donoghue and Menezes \cite{Donoghue:2019ecz,Donoghue:2019fcb,Donoghue:2021eto,Donoghue:2021meq,Donoghue:2021cza} proposed that the higher-derivative ghost can be reinterpreted as a particle traveling back in time, a \textit{Merlin mode}. The Merlin modes are unstable particles and, as such, should not be included in the unitarity relations. Using this property, they were able to prove $S$-matrix unitarity of quadratic gravity \cite{Donoghue:2019fcb}.
	It is important to mention that both the fakeon and Merlin-mode proposals trade unitarity violation with micro-causality violation \cite{Anselmi:2018bra,Anselmi:2018tmf,Donoghue:2019ecz,Donoghue:2021meq,Donoghue:2021cza}. The physical consequences of micro-causality violation need to be better explored. 
	
	Another approach, by Holdom and Ren \cite{Holdom:2015kbf,Holdom:2016xfn,Holdom:2019ouz}, is based on a possible analogy between quantum gravity and quantum chromodynamics (QCD). To our better understanding \cite{Marciano:1977su,Brambilla:2014jmp}, QCD involves different phases: i) at high-energies, QCD is formulated in terms of Yang-Mills theories with non-Abelian gauge fields (\textit{gluons}) coupled to fermions (\textit{quarks}). This formulation enjoys the nice property of asymptotic freedom, with gluons and quarks behaving as free particles in the deep UV-limit \cite{Gross:1973id,Politzer:1974fr}. As we flow towards low energies, gluons and quarks become strongly correlated, and, around the so-called QCD scale, the theory transitions into a second phase. ii) at low energies, QCD can be better described in terms of chiral perturbation theory, where the propagating degrees of freedom are bound states of gluons and quarks \cite{Greensite:2011zz}.
	However, it is still unclear whether this picture is indeed compatible with the dynamics of quadratic gravity. More work in this direction is necessary to make sure that this analogy is indeed viable.
	
	The approach by Holdom and Ren proposes that a similar picture could describe quantum gravity. In this case, in the high energy regime (beyond the Planck scale), the gravitational dynamics would be described by quadratic gravity, a theory that is both renormalizable and (under certain conditions) asymptotically free. In this trans-Planckian regime, the spectrum of the theory contains both gravitons and ghost-like states.
	In analogy to QCD, it is conceivable that once we flow towards low energies, the dynamics of quadratic gravity would enter a strongly correlated regime. Within Holdom-Ren's scenario, the strongly correlated regime would happen around the Planck scale. Below this scale, the theory would transition to a second phase with effective dynamics where the Einstein-Hilbert is the leading order term \cite{Donoghue:1994dn,Donoghue:2012zc,Donoghue:2015hwa,Burgess:2003jk}.
	Furthermore, it is conjectured that the ghost would be removed from the physical spectrum due to non-perturbative effects. Once again, this is based on the analogy with QCD, where gluons and quarks are removed from the physical spectrum due to confinement.
	
	In this paper, we point out that not all the aspects of an analogy between quantum gravity and QCD are desired from the physical point of view. In QCD, the non-perturbative dynamics favors the formation of bound states, leading to a mechanism of chiral symmetry breaking (\chSB{}) \cite{Klevansky:1992qe,Buballa:2003qv,Alkofer:2000wg,Braun:2011pp,Kondo:1991yk,Gies:2002hq,Gies:2005as,Aguilar:2010cn}. As a consequence of \chSB{}, fermions acquire masses of the same order of magnitude as the scale at which QCD becomes strongly correlated.
	Given that gravity is an attractive interaction, we can expect a similar mechanism in a regime where quantum gravity is strongly correlated. However, if this mechanism is indeed realized as part of a theory of quantum gravity that is strongly correlated around the Planck scale, we expect that all fermions would acquire masses around the Planck mass \cite{Eichhorn:2011pc,Meibohm:2016mkp}. This scenario would be in serious contradiction with our observations at low energies since we observe several fermions with masses much lower than the Planck mass. Therefore, a quantum gravity scenario based on an analogy with QCD must avoid \chSB{} around the Planck scale.
	
	The goal of this paper is to present a first analysis of the mechanism of \chSB{} in the scenario of quadratic gravity proposed by Holdom and Ren. We perform an analysis based on functional renormalization group tools \cite{Wetterich:1989xg,Wetterich:1992yh,Morris:1993qb,Ellwanger:1993mw}, where we investigate the impact of graviton fluctuations on the renormalization group flow of four-fermion interactions \cite{Braun:2011pp}. We find indications that, as long as the curvature squared couplings stay finite, four-fermion couplings are not driven toward criticality. Therefore, despite gravity being attractive, it is possible to avoid \chSB{} even in a strongly correlated regime.
	
	This paper is organized as follows: 
	In Sec. \ref{sec:QuadGrav_Rev}, we review the main aspects of quadratic gravity and its analogy to QCD. 
	In Sec. \ref{sec:setup}, we present our setup of investigation.
	In Sec. \ref{sec:RGeqs}, we report the renormalization group equations obtained from our setup. 
	In Sec. \ref{sec:chSB}, we perform an analysis of the phase structure of four-fermion couplings and discuss its consequence to \chSB{}.
	In Sec. \ref{sec:Conclusions}, we present our conclusion and perspectives.
	
	%%%%%%%%%%%%%%%%%%%%%%%%%%%%%%%%%%%%%%%%%%%%%%%%%%%%%%%%%%%%%%%%%%%%%%%%%%%%%%%%%%%%%%%%%%%%
	\section{Quadratic Gravity and its analogy to strong interactions \label{sec:QuadGrav_Rev}}
	
	We can define quadratic gravity as a theory whose dynamics correspond to Einstein-Hilbert plus curvature squared terms. In terms of an action, we can write
	\begin{equation}\label{eq:quadgravaction}
		S_\text{QG} = \int_x \!\sqrt{\det(g)} \,\, \bigg( - M^2 R  + \frac{1}{2f_2^2} C_{\mu\nu\alpha\beta}^2 
		- \frac{1}{3f_0^2} R^2\bigg)\,,
	\end{equation}
	where $M^2$ is a mass parameter, $f_{2}^2$ and $f_{0}^2$ denote dimensionless couplings, $C_{\mu\nu\alpha\beta}$ is the Weyl tensor and $R$ denotes the curvature scalar. Since the main results of this paper are based on Euclidean methods, from now on we use conventions that are compatible with Euclidean signature. Here, we use the $\{R^2,C_{\mu\nu\alpha\beta}^2\}$-basis to define the action for quadratic gravity. We can always translate to other bases by expressing curvature squared invariants in terms of the Gauss-Bonnet invariant, a boundary term in four dimensions. The most general action for quadratic gravity also admits a cosmological constant term. In this paper, we ignore such a term since it does not play an important role in our results.
	
	In order to apply quantum field theory techniques to \eqref{eq:quadgravaction}, we introduce metric fluctuations $h_{\mu\nu}$ according to the linear split $g_{\mu\nu} = \bar{g}_{\mu\nu}  + h_{\mu\nu}$, where $\bar{g}_{\mu\nu}$ denotes a background metric.
	In section \eqref{sec:setup}, we use over-bars to designate geometrical quantities defined in terms of the background metric.
	In the calculation of the renormalization group flow of four-fermion couplings, we can take $\bar{g}_{\mu\nu}$ as a flat metric without loss of generality.
	
	In the perturbative treatment of quadratic gravity, the mass parameter $M^2$ is usually identified with the Planck mass $M_\text{Pl}^2 \sim G_\text{N}^{-1}$ (where $G_\text{N}$ denotes the Newton constant).
	In the scenario proposed by Holdom and Ren \cite{Holdom:2015kbf,Holdom:2016xfn,Holdom:2019ouz}, the mass parameter $M^2$ is not directly identified with $M_\text{Pl}^2$. In fact, in the analogy between quadratic gravity and QCD, we take $M^2 \to 0$ in the trans-Planckian regime.
	Quadratic gravity with $M^2 \to 0$ was also discussed in the context of Agravity \cite{Salvio:2014soa,Salvio:2017qkx} and conformal gravity (where we also take $f_0^2 \to \infty$) \cite{Mannheim:2011ds}.
	
	The analogy between quadratic gravity and QCD starts from the beta functions for the couplings $f_2^2$ and $f_0^2$. Here, we focus on the one-loop beta functions for quadratic gravity (minimally coupled to $N_\text{v}$-vectors, $N_\text{f}$-Dirac fermions and $N_\text{s}$-scalars), previously computed in \cite{Ohta:2013uca,Holdom:2015kbf,Salvio:2018crh}, namely
	\begin{align}
		\beta_{f_2^2} &= - \frac{1}{16\pi^2} \bigg( \frac{133}{10} + 
		\frac{12 N_\text{v} + 3 N_\text{f} + N_\text{s}}{60} \bigg) \, f_2^4 \,, \label{eq:betas_quadgrav_1a} \\
		%%%%%%%%%%%%%%%%%%%%%%%%%%%%%%%%%%%%%%%
		\beta_{f_0^2} &= \frac{1}{16\pi^2} \bigg( \frac{5}{12} f_0^4 + 5 \, f_0^2 f_2^2 
		+ \frac{10}{3} f_2^4 \bigg) \,. \label{eq:betas_quadgrav_1b}
	\end{align}
	The beta function $\beta_{f_2^2}$ (at one-loop) is strictly negative, implying that the coupling $f_2^2$ features the property of asymptotic freedom, connecting $f_2^2=0$ in the deep UV with positive values of $f_2^2$ in the IR. This situation is similar to QCD, where the beta function for the non-Abelian gauge coupling is negative, leading to asymptotic freedom.
	
	\begin{figure}[t]
		\hspace*{-0.8cm}
		\includegraphics[width=.8\linewidth]{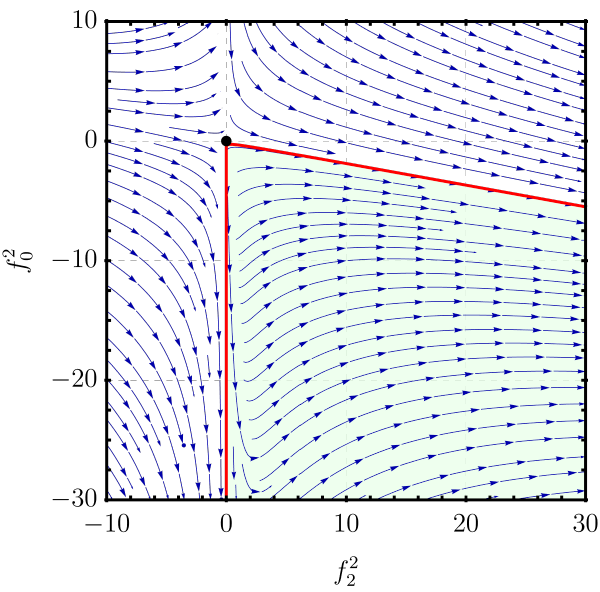}
		\caption{Phase diagram of the gravitational couplings $f_2^2$ and $f_0^2$. The green region corresponds to the basin of attraction of the UV fixed point $(f_{2,*}^2,f_{0,*}^2)=(0,0)$.}
		\label{fig:phasegrav}
	\end{figure}
	The analysis of the beta function $\beta_{f_0^2}$ is a bit more subtle since it does not have a definite sign. For positive values of $f_0^2$ and $f_2^2$, it is clear that $\beta_{f_0^2}$ is positive, indicating that we cannot have asymptotically free trajectories connected to positive values of $f_0^2$ and $f_2^2$ in the IR. If $f_0^2 < 0$ in the infrared, it is possible to find asymptotically free trajectories for $f_0^2$. In Fig. \ref{fig:phasegrav}, we show the phase diagram of the curvature squared couplings.
	In the presence of an Einstein-Hilbert term in \eqref{eq:quadgravaction}, $f_0^2 < 0$ leads to a tachyon in the spectrum obtained from the tree-level propagator. Here, we disregard such an issue, since we will work in a (conjectured) setting where the tree-level propagator does not capture the actual physical states of the theory. 
	
	In the scenario proposed by Holdom and Ren \cite{Holdom:2015kbf,Holdom:2016xfn,Holdom:2019ouz}, they conjectured that the analogy between quadratic gravity and QCD goes beyond the running of $f_2^2$ and $f_0^2$.
	The main conjecture is that the dressed propagator associated with metric fluctuations should inherit certain aspects of the dressed gluon propagator.
	In QCD, the tree-level gluon propagator reads (apart from tensorial structures)
	\begin{equation}
		\mathcal{D}_{AA}^{\text{tree}}(q^2) \sim \frac{1}{q^2} \,.
	\end{equation}
	The tree-level gluon propagator suggests that QCD describes colored massless particles as asymptotic states. However, this picture seems to be in contradiction with color confinement.
	The true spectrum of QCD should be obtained from dressed propagators, where we expect that non-perturbative effects will properly incorporate color confinement. In general, we can write the dressed gluon propagator as
	\begin{equation}
		\mathcal{D}_{AA}(q^2) \sim \frac{F_{AA}(q^2)}{q^2} \,,
	\end{equation}
	where the form factor $F_{AA}(q^2)$ contains the non-perturbative information.
	In the last two decades, the QCD community has made significant progress in obtaining non-perturbative information about the gluon propagator from various methods, including lattice simulations and functional methods \cite{Cucchieri:2007md,Cucchieri:2007rg,Oliveira:2012eh,Maas:2011se,Aguilar:2008xm,Fischer:2008uz,vonSmekal:1997ohs,Dudal:2008sp,Boucaud:2008ky}. The different approaches to non-perturbative QCD agree that the gluon propagator (at least in the Landau gauge) features a decoupling regime with $\mathcal{D}_{AA}(0)$ approaching a positive and finite value. Lattice simulations indicate that one can fit the  gluon propagator (modulo logarithmic corrections) by a four-parameter function, producing the form factor 
	\begin{equation}
		F_{AA}^\text{fit}(q^2) = \frac{Z \,q^2\,(q^2 + M_1^2)}{q^4 + M_2^2\,q^2 + M_3^4} \,.
	\end{equation}
	The fitting obtained with quenched SU(3) gauge fields indicates that the dressed gluon propagator has only complex poles \cite{Cucchieri:2011ig,Dudal:2010tf}. Thus, it is not physically meaningful to associate a particle interpretation to the poles of the dressed gluon propagator. A possible interpretation is that one can associate complex poles in the gluon propagator with color confinement \cite{Hayashi:2020few}.
	
	Now, we turn to propagators in quadratic gravity. Keeping aside tensorial structures, the tree-level propagator associated with metric fluctuations behaves as
	\begin{equation}
		\mathcal{D}_{hh}^{\text{tree}}(q^2) \sim \frac{1}{q^4} \,,
	\end{equation}
	This type of behavior is problematic since it is difficult to reconcile $1/q^{4}$-scaling with unitarity. Holdom and Ren suggested that the tree-level propagator in quadratic gravity is not the appropriate quantity to define asymptotic states in quantum gravity. Instead, we should look at the dressed propagator, which we parameterize as
	\begin{equation}
		\mathcal{D}_{hh}(q^2) \sim \frac{F_{hh}(q^2)}{q^4} \,,
	\end{equation}
	with $F_{hh}(q^2)$ representing a form factor. The conjecture by Holdom and Ren is that, at the non-perturbative level, the form factor $F_{hh}(q^2)$ behaves similarly to $F_{AA}(q^2)$. If this conjecture is correct, we can write
	\begin{equation}
		\mathcal{D}_{hh}(q^2) \sim \frac{1}{q^2} \,\mathcal{D}_{AA}(q^2)\, .
	\end{equation}
	Thus, keeping in mind that the dressed gluon propagator features only complex poles, the only real pole of $\mathcal{D}_{hh}(q^2)$ is the one at $q^2 = 0$. This massless pole is what one can interpret as the graviton, and this would be the only asymptotic state in quadratic gravity. In this picture, the original poles of the tree-level propagator $\mathcal{D}_{hh}^{\text{tree}}$ would be removed from the physical spectrum.
	
	To complete the scenario, Holdom and Ren conjecture that once gravity enters a strongly correlated regime, it will undergo a phase transition, such that it would be described by an effective field theory below the Planck, where the massless graviton is the only propagating degree of freedom.
	Again, this is based on the analogy with QCD. The physics of strong interactions is well described by chiral perturbation theory for energies below the QCD phase transition. 
	In contrast with QCD, the effective theory of quantum gravity below the Planck scale would still involve metric fluctuations as its dynamical variable. However, the dynamics below the Planck scale would be dictated by the Einstein-Hilbert action plus corrections suppressed by negative powers of the Planck mass.
	
	While the scenario reviewed in this section is a conjecture, we can already ask some questions about it. In particular, what aspects of QCD should be imported to quantum gravity in this analogy? While the analogy at the level of the dressed propagator is very welcome, there are aspects of QCD that one needs to avoid in a theory of quantum gravity that enters a non-perturbative regime around the Planck scale.
	In QCD, \chSB{} is the leading order mechanism behind the generation of the mass of hadrons.
	This mechanism relies on two main aspects: i) quarks are attracted to each other due to the exchange of gluons, which favors bound-state formation; ii) the non-Abelian interaction enters a strongly correlated regime around the QCD scale $\Lambda_\text{QCD} \sim 10^2 \,\text{MeV}$. As a consequence of \chSB{}, bound states of colored fermions acquire masses of the same order as $\Lambda_\text{QCD}$.
	
	In the scenario discussed in this paper, gravity fulfills the same conditions that are relevant for \chSB{}: i) gravity is an attractive interaction; ii) quadratic gravity enters a strongly correlated regime around the Planck scale. Thus, it is reasonable to expect that quantum gravity effects can trigger \chSB{} around the Planck scale. If this is the case, we would expect that all fermions in nature (as gravity is universal) should form bound states with masses around the Planck scale. This picture is in clear contradiction with our Universe since we observe fermions that are much lighter than the Planck mass. 
	In the rest of this paper, we test whether the scenario proposed by Holdom and Ren leads to \chSB{}.
	
	%%%%%%%%%%%%%%%%%%%%%%%%%%%%%%%%%%%%%%%%%%%%%%%%%%%%%%%%%%%%%%%%%%%%%%%%%%%%%%%%%%%%%%%%%%%%
	\section{Technical setup \label{sec:setup}}
	
	The main goal of this paper is to investigate if quantum gravity fluctuations would trigger \chSB{} in the scenario of quadratic gravity in analogy to QCD.
	One efficient way of searching for \chSB{} is by looking at the renormalization group flow of four-fermion interactions. See, \textit{e.g.}, \cite{Braun:2011pp} for a review on the role of four-fermion interactions in the mechanism of \chSB{}. In a nutshell, one can interpret divergences in the flow of four-fermion couplings as indications for \chSB{}.
	
	In this paper, we use the functional renormalization group (FRG) as a tool to derive beta functions in fermionic systems coupled to quadratic gravity.
	The FRG is a practical implementation of the Wilsonian idea of defining the renormalization group flow by a step-by-step realization of (Euclidean) functional integrals in quantum field theory \cite{Wetterich:1989xg,Wetterich:1992yh,Morris:1993qb,Ellwanger:1993mw}.
	Within the \textit{effective average action} formulation, the main idea is to deform the generating functional by introducing a regulator function $\textbf{R}_k(q^2)$ that acts like a momentum-dependent mass term, with $k$ denoting a renormalization group scale.
	By construction, $\textbf{R}_k(q^2)$ should: i) suppress IR modes characterized by $q^2 < k^2$; ii) vanish for $k \to 0$; iii) diverge when $k \to \infty$; iv) approach zero in a sufficiently quick way when we set $q^2 \to \infty$ with fixed $k$.
	For reviews on the FRG, see, \textit{e.g.}, Refs. \cite{Pawlowski:2005xe,Gies:2005as,Rosten:2010vm,Dupuis:2020fhh}. For applications of the FRG in quantum gravity, see, \textit{e.g.}, Refs. \cite{Reuter:1996cp,Percacci:2017fkn,Reuter:2019byg,Dupuis:2020fhh}.
	
	The main object of the FRG is the effective average action $\Gamma_k$, which is a functional that describes the effective dynamics of a system at scale $k$, recovering the full effective action $\Gamma$ at $k=0$ and the bare action at $k\to \Lambda_\text{UV}$ ($\Lambda_\text{UV}$ is a UV cutoff, we can set it to infinity in UV-complete theories).
	The functional $\Gamma_k$ satisfies the non-perturbative flow equation,
	\begin{equation}\label{eq:floweq}
		k \,\pt_k \Gamma_k = \frac{1}{2} \STr \Big[ \Big( \Gamma^{(2)}_k + \textbf{R}_k \Big)^{-1} \, k\, \pt_k \textbf{R}_k \Big] \,,
	\end{equation}
	where $\Gamma^{(2)}_k$ is a 2-point function derived from $\Gamma_k$, and the super-trace $\STr$ accounts for functional and discrete traces, and also for negative signs in the case of fermions.
	
	The flow equation \eqref{eq:floweq} is a powerful tool to derive beta functions beyond standard perturbative approximations. However, we cannot solve this equation exactly. We still need to employ some form of approximation method. Typically, we start with an \textit{ansatz}, \textit{i.e.}, a truncation for $\Gamma_k$, then we compute beta functions within the corresponding truncated coupling space.
	
	In this paper, we use the following truncation for $\Gamma_k$
	\begin{equation}\label{eq:truncation}
		\begin{aligned}
			&\Gamma_k =
			\int_x \!\sqrt{\det(g)} \,\, \bigg(  \frac{1}{2f_2^2(k)} C_{\mu\nu\alpha\beta}^2 
			- \frac{1}{3f_0^2(k)} R^2\bigg) \\
			&\,\,\,+\int_x \! \sqrt{\det(\bar{g})} \,
			\, F_\mu[h,\bar{g}] Y^{\mu\nu}(\bar{g}) F_\nu [h,\bar{g}] \\
			&\,\,\,+ \int_x \!\sqrt{\det(g)} \,\, i \, Z_\psi(k)\, \bar{\psi}_i \, \gamma^\mu D_\mu \psi_i \\
			&\,\,\,+\int_x \!\sqrt{\det(g)} \left( \frac{\bar{\lambda}_+(k)}{2} \big( \text{V} + \text{A} \big) 
			+ \frac{\bar{\lambda}_-(k)}{2} \big( \text{V} - \text{A} \big)\right)  \,.
		\end{aligned}
	\end{equation}
	The first and second lines correspond to the truncation of the gravitational dynamics. We use a scale-dependent version of the quadratic gravity action (with $M^2 \to 0$), plus a higher-derivative gauge-fixing term with \cite{Barth:1983}
	\begin{align}
		&F_\mu[h,\bar{g}] = \frac{1}{\sqrt{2\, \alpha_\text{gf}}} \left( \bar{\nabla}^\nu h_{\mu\nu} - \frac{1+\beta_\text{gf}}{4} \nabla_\mu h^\alpha_{\,\,\,\alpha} \right) \,, \\
		&Y^{\mu\nu}(\bar{g}) = -\, \bar{g}_{\mu\nu} \bar{\nabla}^2 + c_\text{gf} \bar{\nabla}_\mu\bar{\nabla}_\nu - d_\text{gf} \bar{\nabla}_\nu\bar{\nabla}_\mu \,.
	\end{align}
	Here, $\alpha_{\text{gf}}$, $\beta_{\text{gf}}$, $c_{\text{gf}}$ and $d_{\text{gf}}$ denote gauge parameters. In this paper, we report results obtained with gauge choice $\alpha_{\text{gf}} = f_2^2$, $\beta_{\text{gf}} = 3 f_2^2/(f_2^2 + f_0^2)$ and $c_{\text{gf}} - d_{\text{gf}} = ( f_0^2 - 2f_2^2)/(3f_0^2)$. This gauge choice is adjusted to remove off-diagonal terms in the graviton 2-point function \cite{Barth:1983,Codello:2006in,Ohta:2013uca}.\\
	The third and fourth lines of \eqref{eq:truncation} correspond to the truncation of the fermion dynamics. Our truncation for the fermion sector corresponds to a Nambu-Jona-Lasino \cite{Nambu:1961tp,Nambu:1961fr} type of action minimally coupled to gravity (via vierbein formalism). In the fourth line, $\text{V}$ and $\text{A}$ denote, respectively, the vector and axial-vector four-fermion operators, namely
	\begin{align}
		\text{V} = (\bar{\psi}_i \gamma_\mu \psi_i)^2
		\quad \text{and} \quad
		\text{A} = -(\bar{\psi}_i \gamma_\mu \gamma_5 \psi_i)^2 \,.
	\end{align}
	The index $i$ represents a flavor-index, taking values $i \in \{1,\dots, N_\text{f}\}$. 
	In the fermionic sector, we introduce a wave-function renormalization $Z_\psi (k)$ and two four-fermion couplings $\bar{\lambda}_+(k)$ and $\bar{\lambda}_-(k)$ (with canonical mass dimention $-2$).
	In order to derive beta functions that lead to autonomous dynamical systems, we define the renormalized four-fermion couplings
	\begin{equation}
		\lambda_\pm(k) = \big(Z_\psi(k)^{2} \, k^{-2} \big)^{-1}\, \bar{\lambda}_\pm(k) \,.
	\end{equation}
	
	The system in \eqref{eq:truncation} is symmetric under the global chiral group $\text{SU}(N_\text{f})_\text{R} \times \text{SU}(N_\text{f})_\text{R}$, corresponding to transformations of the form
	\begin{align}
		&\psi_i \mapsto \psi'_i = (U_\text{R})_{ij} \, P_\text{R} \psi_j + (U_\text{L})_{ij} \, P_\text{L} \psi_j \,,\\
		&\bar{\psi}_i \mapsto \bar{\psi}'_i = \bar{\psi}_{j} P_\text{L} \, (U_\text{R}^\dagger)_{ji}  
		+ \bar{\psi}_{j} P_\text{R} \, (U_\text{L}^\dagger)_{ji} \,,
	\end{align}
	where $U_\text{R} \in \text{SU}(N_\text{f})_\text{R}$ and $U_\text{L} \in \text{SU}(N_\text{f})_\text{L}$, and $P_{\text{R},\text{L}}=\frac{1}{2}(\mathbf{1}\pm \gamma_5)$ are the chiral projectors.
	In addition, \eqref{eq:truncation} is also symmetric under the global chiral transformation
	\begin{equation}
		\psi_i \mapsto \psi'_i = e^{i \alpha_i \gamma_5} \,\psi_i
		\quad \text{and} \quad
		\bar{\psi}_i \mapsto \bar{\psi}'_i = \bar{\psi}_i \,e^{i \alpha_i \gamma_5},
	\end{equation}
	for each one of the flavors.
	
	For practical calculations with the FRG, we need to specify the regulator function $R_k(q^2)$. In this paper, we define
	\begin{eqnarray}
		\textbf{R}_k(q^2) = [\Gamma_k^{(2)}(q^2)]_{\text{fields}=0} \,\, r(q^2) \,,
	\end{eqnarray}
	where we use $r(q^2) = \big((k^2/q^2)^2 - 1 \big) \, \theta(k^2- q^2)$ in the gravitational sector, and $r(q^2) = \big( (k^2/q^2)^{1/2}-1 \big) \, \theta(k^2- q^2)$ in the fermionic sector.
	%%%%%%%%%%%%%%%%%%%%%%%%%%%%%%%%%%%%%%%%%%%%%%%%%%%%%%%%%%%%%%%%%%%%%%%%%%%%%%%%%%%%%%%%%%%%
	\section{Renormalization Group Equations  \label{sec:RGeqs}}
	
	Using the FRG, one can compute the beta-functions $\beta_{\lambda_+}$ and $\beta_{\lambda_-}$ by acting with functional derivatives w.r.t. fermion on both sides of the flow equation \eqref{eq:floweq}, and projecting it at vanishing fields. Schematically, we can write
	\begin{equation}
		\beta_{\lambda_\pm} = (2 + 2 \,\eta_\psi) \, \lambda_\pm \,+\, 
		\mathcal{P}_\pm \circ k\pt_k \Gamma^{(4)}_{k,\psi\psi\bar{\psi}\bar{\psi}} \,,
	\end{equation}
	where $\eta_\psi = - Z_\psi^{-1} \, k \pt_k Z_\psi$ is the fermion anomalous dimension, $\mathcal{P}_+$ and $\mathcal{P}_-$ are operators that allow us to project on $\lambda_+$ and $\lambda_-$, and $\Gamma^{(4)}_{k,\psi\psi\bar{\psi}\bar{\psi}}$ is the fermion 4-point function evaluated at vanishing momenta. 
	At the technical level, we use the \textit{Mathematica} packages \textit{xAct} \cite{xAct1,xAct2,xAct3}, \textit{FormTracer} \cite{FormTracer} and \textit{DoFun} \cite{DoFun1,DoFun2} to evaluate the beta functions $\beta_{\lambda_\pm}$.
	
	In the following, we report the explicit expressions for $\beta_{\lambda_+}$ and $\beta_{\lambda_-}$ in the perturbative approximation, where we ignore the anomalous dimension $\eta_\psi$ coming from the regulator insertion $k\pt_k \textbf{R}_k$,
	\begin{equation}
		\begin{aligned}\label{eq:betas_four-fermion}
			\beta_{\lambda_\pm} &= \beta_{\lambda_\pm}^{(0)} 
			\,\pm\, \frac{f_2^2}{28672\,\pi^2} \, \left( 141 f_2^2 - \frac{811}{55}\, \beta_{f_2^2} \right)  \\
			& - \frac{\lambda_\pm}{192\pi^4} \, \bigg( 67\, f_2^2  - \frac{779}{120} \beta_{f_2^2}
			- 8\,f_0^2 + \frac{23}{60} \beta_{f_0^2} \bigg) \,,
		\end{aligned}
	\end{equation}
	where $\beta_{\lambda_+}^{(0)}$ and $\beta_{\lambda_-}^{(0)}$ denote the beta functions of four-fermion couplings in the absence of gravity, namely
	\begin{align}
		\beta_{\lambda_+}^{(0)} &= 2 \lambda_+ + \frac{3}{8\pi^2} \,\lambda_+^2 + \frac{N_\text{f}+1}{4\pi^2}\, \lambda_+ \lambda_- \,, \\
		\beta_{\lambda_-}^{(0)} &= 2 \lambda_- + \frac{N_\text{f} - 1}{8\pi^2} \,\lambda_-^2 + \frac{N_\text{f}}{8\pi^2} \,\lambda_+^2 \,.
	\end{align}
	The beta functions $\beta_{\lambda_\pm}^{(0)}$ agree with the  limit of vanishing gauge couplings of the results reported in \cite{Gies:2002hq,Gies:2003dp,Gies:2005as,Braun:2006jd}. 
	
	Concerning the renormalization group flow of the gravitational parameters $f_2^2$ and $f_0^2$, we use the 1-loop beta functions previously computed in the literature (c.f.~Eq.~ \eqref{eq:betas_quadgrav_1a} and \eqref{eq:betas_quadgrav_1b}) \cite{Ohta:2013uca,Holdom:2015kbf,Salvio:2018crh}. Since we are interested in a system containing only gravity and fermions, we set $N_\text{v} = N_\text{s} = 0$, resulting in
	\begin{align}
		\beta_{f_2^2} &= - \frac{1}{16\pi^2} \bigg( \frac{133}{10} + 
		\frac{ N_\text{f} }{20} \bigg) \, f_2^4 \,, \label{eq:betas_quadgrav_2a} \\
		%%%%%%%%%%%%%%%%%%%%%%%%%%%%%%%%%%%%%%%
		\beta_{f_0^2} &= \frac{1}{16\pi^2} \bigg( \frac{5}{12} f_0^4 + 5 \, f_0^2 f_2^2 
		+ \frac{10}{3} f_2^4 \bigg) \,. \label{eq:betas_quadgrav_2b}
	\end{align}
	Ideally, we should not use 1-loop beta functions to explore the strongly correlated regime. This is a limitation of our current analysis. 
	In Sec. \ref{sec:chSB}, we explore a toy model that parameterizes the renormalization group flow of the gravitational couplings in the strongly correlated regime.
	%%%%%%%%%%%%%%%%%%%%%%%%%%%%%%%%%%%%%%%%%%%%%%%%%%%%%%%%%%%%%%%%%%%%%%%%%%%%%%%%%%%%%%%%%%%%
	\section{\chSB{} in quadratic gravity? \label{sec:chSB}}
	
	In this section, we search for signs of \chSB{} by looking at the renormalization group flow of the four-fermion couplings $\lambda_+$ and $\lambda_-$.
	
	\subsection{Switching off gravitation fluctuations}
	
	First, we note that if we switch off gravitational fluctuations ($f_2^2\to0$ and $f_0^2\to0$), $\beta_{\lambda_\pm}$ reduces to $\beta_{\lambda_\pm}^{(0)}$.
	The beta functions $\beta_{\lambda_\pm}^{(0)}$ feature one free fixed point,
	\begin{equation}\label{eq:fixpt_PMfree}
		(\lambda_{+,*},\lambda_{-,*}) = (0,0) \,,
	\end{equation}
	and three interacting fixed points,
	\begin{align}\label{eq:fixpt_PMint}
		\hspace*{-.18cm}(\lambda_{+,*},\lambda_{-,*}) &= \left(-\frac{16\pi^2 \,(3+N_\text{f})}{9 + 5N_\text{f} + 2N_\text{f}^2} \, , \, -\frac{16\pi^2}{9 + 5N_\text{f} + 2N_\text{f}^2} \right)  \,, \\
		\hspace*{-.18cm}(\lambda_{+,*},\lambda_{-,*}) &= \left( -\frac{16\pi^2}{1-2N_\text{f}} \, , \, \frac{16\pi^2}{1-2N_\text{f}} \right) \,,\\
		\hspace*{-.18cm}(\lambda_{+,*},\lambda_{-,*}) &= \left( 0 \, , \, \frac{16\pi^2}{1-N_\text{f}} \right) \,.
	\end{align}
	See also Ref.~\cite{Braun:2011pp} for a discussion on the fixed-point structure of $\beta_{\lambda_\pm}^{(0)}$.
	Here, we focus on the case $N_\text{f}\geq2$. We will discuss the single-favor ($N_\text{f}=1$) case in section \ref{sec:chSBSingleFlavor}.
	
\begin{figure}[t]
		\hspace*{-0.8cm}
		\includegraphics[width=.8\linewidth]{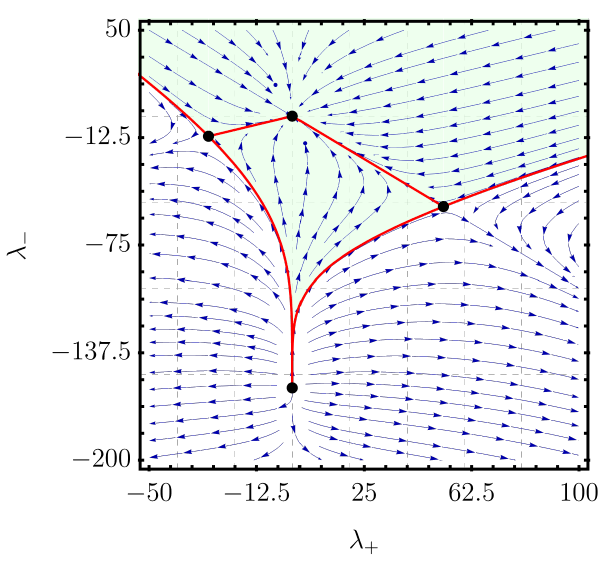}
		\caption{
			Phase diagram of four-fermion interaction with $N_\text{f}=2$ and in the absence of gravitational contributions. 
			The green region indicates the chiral phase. Any UV-initial condition inside this region leads to trajectories that are attracted to the IR fixed point $\text{FP}_1^{(0)}$. 
			The red lines correspond to separatrix between the different sub-regions of the phase diagram.}
		\label{fig:phase_nograv_Nf=2}
	\end{figure}

	We label the fixed points as $\textbf{\text{FP}}_1^{(0)}$, $\textbf{\text{FP}}_2^{(0)}$, $\textbf{\text{FP}}_3^{(0)}$ and $\textbf{\text{FP}}_4^{(0)}$, respectively.
	Concerning their stability properties, we have that: 
	\begin{itemize}
		\item $\textbf{\text{FP}}_1^{(0)}$ is IR-attractive;
		\item $\textbf{\text{FP}}_2^{(0)}$ and $\textbf{\text{FP}}_3^{(0)}$ have one UV-attractive and one IR-attractive direction; 
		\item $\textbf{\text{FP}}_4^{(0)}$ is UV-attractive.
	\end{itemize}
	This set of fixed points defines a region (c.f.~Fig.~\ref{fig:phase_nograv_Nf=2}) in the phase space of four-fermion couplings for which any initial condition inside such region leads to safe trajectories towards the IR, thus, avoiding \chSB{}. We will use the nomenclature \textit{chiral phase} to refer to this region.
	
	When we switch on gravitational contributions in the beta functions $\beta_{\lambda_\pm}$, the chiral phase will be deformed due to non-vanishing values of $f_2^2$ and $f_0^2$. Hence, we can search for signs of gravity-induced \chSB{} by investigating deformations of the chiral phase for non-vanishing values of $f_2^2$ and $f_0^2$. In particular, if the chiral phase gets shrunk to a point for finite values of $f_2^2$ and $f_0^2$, we can view this as an indication that quadratic gravity would trigger \chSB{} if it enters a non-perturbative regime.
	
	\subsection{A brief detour into single-channel QCD}
	\begin{figure*}[t!]
		\hspace*{-0.8cm}
		\includegraphics[width=.9\linewidth]{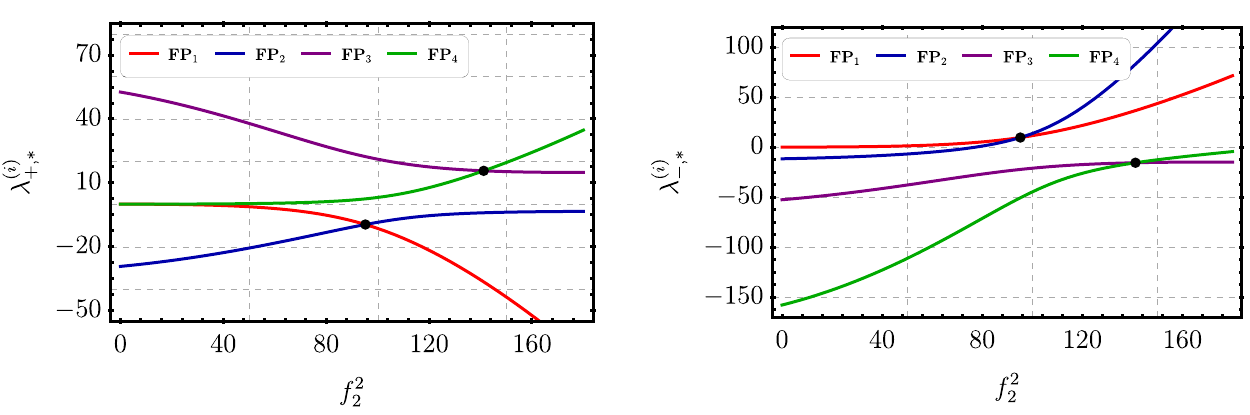}
		\caption{
			We show the fixed-point values $\lambda_{+,*}$ and $\lambda_{-,*}$ as a function of $f_2^2$ in the TT-approximation.
			Different lines correspond to the different fixed points $\text{\textbf{FP}}_i$ with $i\in \{1,2,3,4\}$. This plot corresponds to $N_\text{f}=2$.
			The black dots indicate fixed-point collisions.	
		}
		\label{fig:lambdafp _TTapprox}
	\end{figure*}
	
	To better illustrate our strategy, let us briefly review how to identify \chSB{} in QCD with a simple setup where we consider only the $(\text{S} - \text{P})$-channel, associated with a four-fermion coupling $\lambda_\sigma$. Our discussion follows \cite{Gies:2002hq,Gies:2003dp,Gies:2005as,Braun:2006jd,Braun:2011pp}. In this single-channel approximation, we search for \chSB{} by looking at zeros of the beta function
	\begin{equation}
		\beta_{\lambda_\sigma}^\text{QCD} = 2 \,\lambda_\sigma
		- \frac{3}{4\pi^2} \, \lambda_\sigma^2 - \frac{1}{\pi^2} g^2 \,\lambda_\sigma 
		- \frac{57}{256\pi^2} g^4 \,,
	\end{equation}
	where $g$ denotes the SU(3) gauge coupling. In the following, we will treat $g$ as an external parameter.
	One can define the chiral phase as the region between the fixed points (zeros of the beta function) $\lambda_{\sigma*}^\text{(IR)}$ and $\lambda_{\sigma,*}^\text{(UV)}$, where
	\begin{align}
		\!\!\!\lambda_{\sigma,*}^\text{(IR)} &= 
		\frac{32\pi^2 \!-\! 16 g^2 \!-\! \sqrt{1024\pi^4 - 1024\pi^2 \, g^2 + 85\,g^4}}{24} \,,\\
		\!\!\!\lambda_{\sigma,*}^\text{(UV)} &= 
		\frac{32\pi^2 \!-\! 16 g^2 \!+\! \sqrt{1024\pi^4 - 1024\pi^2 \, g^2 + 85\,g^4}}{24} \,.
	\end{align}
	If $\lambda_{\sigma,*}^\text{(IR)}, \lambda_{\sigma,*}^\text{(UV)} \in \mathbb{R}$, then any initial condition in the chiral phase flows towards $\lambda_{\sigma,*}^\text{(IR)}$ in the IR.
	However, we note that the chiral phase exists only if the argument of the square roots in the previous expressions remains positive. Thus, the requirement of a non-vanishing chiral phase leads to an upper bound $g^2 < g^2_\text{crit} \approx 10.86 $. 
	At $g^2 = g^2_\text{crit}$ we have $\lambda_{\sigma,*}^\text{(IR)} = \lambda_{\sigma,*}^\text{(UV)}$, corresponding to a \textit{fixed-point collision}.
	For $g^2>g^2_\text{crit}$, $\lambda_{\sigma,*}^\text{(IR)}$ and $\lambda_{\sigma,*}^\text{(UV)}$ runs into the complex plane, and the flow of the four-fermion coupling $\lambda_\sigma$ becomes unbounded.
	Therefore, if the flow of $g^2$ stays long enough above $g^2_\text{crit}$, then the four-fermion coupling $\lambda_\sigma$  runs towards a divergence, leading to \chSB{}.
	
	%%%%%%%%%%%%%%%%%%%%%%%%%%%%%%%%%%%%%%%%%%%%%%%%%%%%%%%%%%%%%%%%%%%%%%%
	%%%%%%%%%%%%%%%%%%%%%%%%%%%%%%%%%%%%%%%%%%%%%%%%%%%%%%%%%%%%%%%%%%%%%%%
	%%%%%%%%%%%%%%%%%%%%%%%%%%%%%%%%%%%%%%%%%%%%%%%%%%%%%%%%%%%%%%%%%%%%%%%
	\subsection{\chSB{} in Quadratic Gravity: $N_\text{f} \geq 2$}
	\begin{figure*}[t!]
		\includegraphics[width=.85\linewidth]{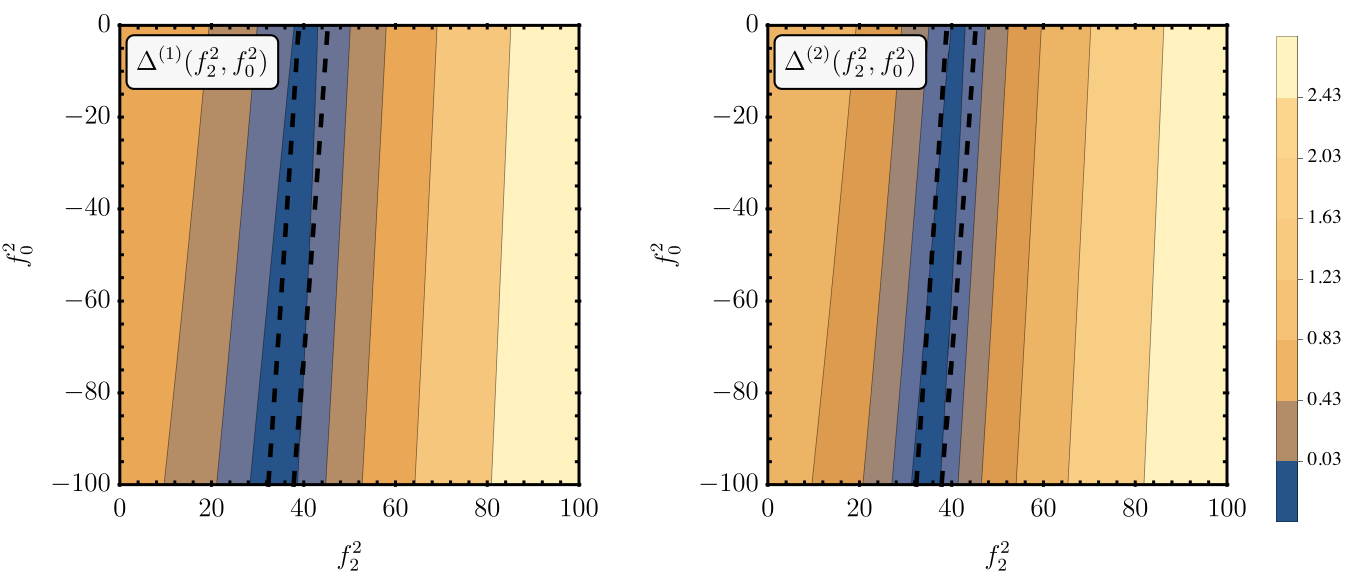}
		\caption{
			Contour plots showing how the quantities $\Delta^{(1)}$ and $\Delta^{(2)}$ (see Eq. \eqref{eq:Deltas}) vary with respect to the gravitational couplings $f_2^2$ and $f_0^2$. 
			This plot corresponds to $N_f=2$.
			As we can see, $\Delta^{(1)}$ and $\Delta^{(2)}$ is mainly affected by changes on $f_2^2$.
			The dashed lines indicate fixed-point collisions on the $(f_2^2,f_0^2)$-plane.
			We define: i) $\mathcal{R}_1$ as the region to the left of the first dashed line; ii) $\mathcal{R}_2$  as the region between the dashed lines; iii) $\mathcal{R}_3$  as the region to the right of the second dashed line.
		}
		\label{fig:DeltaContour_Nf2}
	\end{figure*}	
	\begin{figure*}[t!]
		\hspace*{-0.8cm}
		\includegraphics[width=.9\linewidth]{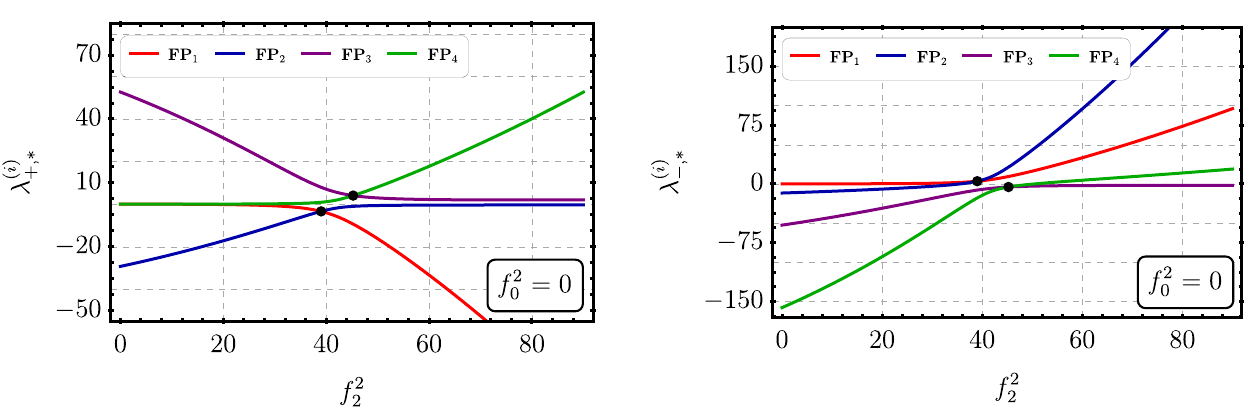}
		\caption{
			We show the fixed-point values $\lambda_{+,*}$ and $\lambda_{-,*}$ as a function of $f_2^2$.
			Different lines correspond to the different fixed points $\text{\textbf{FP}}_i$ with $i\in \{1,2,3,4\}$. In this plot, we use $N_\text{f}=2$ and $f_0^2 = 0$ as a representative case.
			The black dots indicate fixed-point collisions.}
		\label{fig:lambda_fp}
	\end{figure*}	
	\begin{figure*}[t!]
		\includegraphics[width=\linewidth]{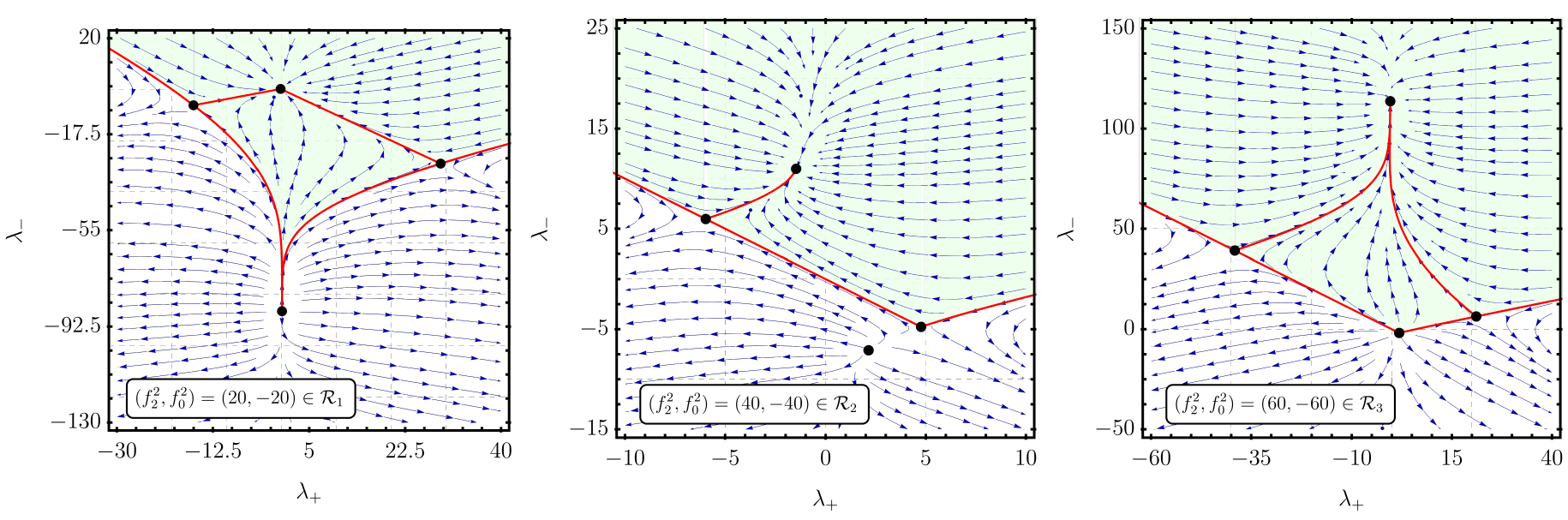}
		\caption{
			Phase diagram of four-fermion interaction including gravitational contributions. 
			The green region indicates the chiral phase. In all plots, we show the results corresponding to $N_\text{f}=2$.}
		\label{fig:phase_withgrav_Nf=2}
	\end{figure*}
	\begin{figure*}[t]
		\hspace*{-0.8cm}
		\includegraphics[width=.85\linewidth]{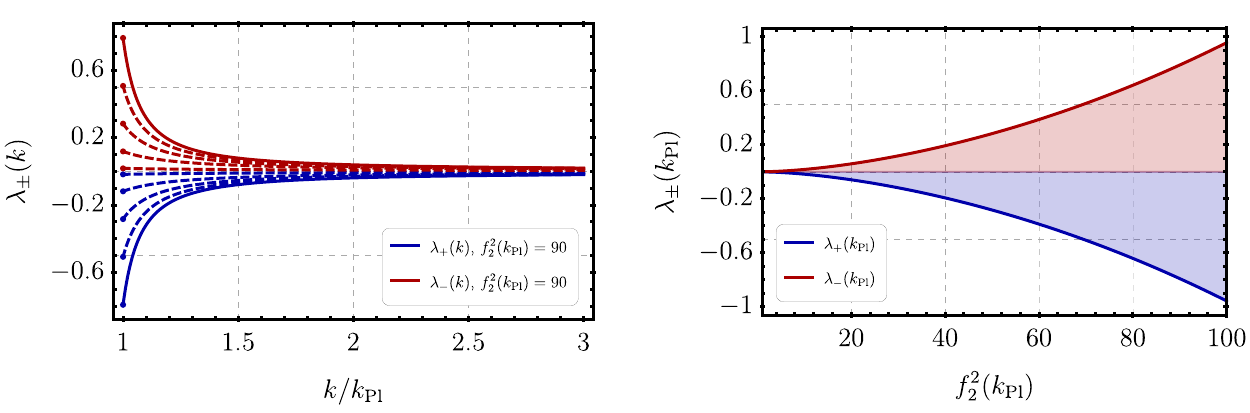}
		\caption{
			Left panel: Renormalization group trajectories of the four-fermion couplings $\lambda_{\pm}$. We show trajectories corresponding to different boundary conditions for the gravitational couplings at the Planck scale.
			Right panel: Planck scale value of $\lambda_\pm$ as a function of the gravitational $f_2^2(k_\text{Pl})$.
			In all cases, with focus on $f_0^2(k_\text{Pl}) = - f_2^2(k_\text{Pl})$ and $N_\text{f}=2$
		}
		\label{fig:lambdatraject}
	\end{figure*}
	\begin{figure*}[t]
		\hspace*{-0.8cm}
		\includegraphics[width=.85\linewidth]{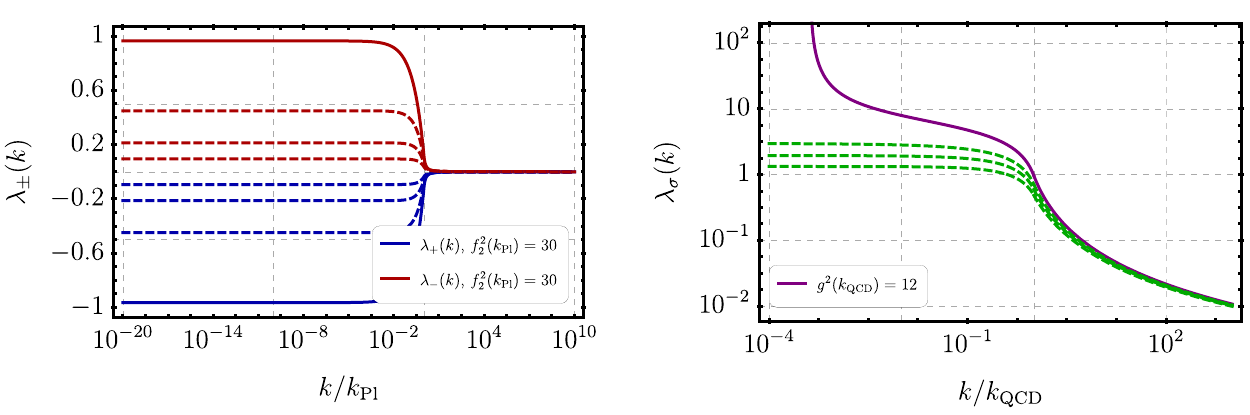}
		\caption{Left panel: Renormalization group trajectories of the four-fermion couplings $\lambda_{\pm}$ integrated down to IR-scales. The external trajectories (full lines) correspond to the boundary condition $f_2^2(k_\text{Pl}) = -f_0^2(k_\text{Pl}) = 30$.
			For the internal trajectories (dashed lines) we use boundary conditions $f_2^2(k_\text{Pl}) = -f_0^2(k_\text{Pl}) \in \{15,20,25\}$.
			The four-fermion couplings remain finite even if we choose large boundary conditions at the Planck scale $k_\text{Pl}$.
			Right panel: Renormalization group trajectories of four-fermion coupling $\lambda_\sigma$ in our single-channel toy-model for QCD. 
			The full line corresponds to boundary condition $g^2(k_\text{QCD}) = 12$. The dashed lines correspond to boundary conditions $g^2(k_\text{QCD}) \in \{8,9,11\}$.
			In the QCD example, $\lambda_\sigma(k)$ diverges if we choose large enough boundary conditions for the gauge coupling at the transition scale $k_\text{QCD}$.}
		\label{fig:lambdaGravQCD}
	\end{figure*}

	Next, we investigate if a similar situation happens in the context of quadratic gravity.
	First, we look at a simplified setting where we consider only gravitational contributions associated with the \textit{transverse and traceless} modes of the fluctuation field $h_{\mu\nu}$, \textit{i.e.}, the TT-approximation. In this case, the beta functions of $\lambda_\pm$ reduce to
	\begin{equation}
		\begin{aligned}
			\beta_{\lambda_\pm}\big|_\text{TT} &= \beta_{\lambda_\pm}^{(0)} 
			\,\pm\, \frac{f_2^2}{2048\,\pi^2} \, \left( 10 f_2^2 - \, \beta_{f_2^2} \right)  \\
			& - \frac{5\,\lambda_\pm}{64\pi^4} \, \bigg( f_2^2  - \frac{1}{8} \beta_{f_2^2} \bigg) \,,
		\end{aligned}
	\end{equation}
	Treating $f_2^2$ as an external parameter, this system of beta functions has four fixed points with the following structure
	\begin{equation}\label{eq:struct_zeros}
		\textbf{\text{FP}}_i: \quad \lambda_{\pm,*}^{(i)} = A_\pm^{(i)} + B_\pm^{(i)} \, \sqrt{\Delta^{(i)}} \,,
	\end{equation}
	where $i \in \{1,2,3,4\}$. 
	We sorted the fixed points such that $\textbf{\text{FP}}_i \to \textbf{\text{FP}}_i^{(0)}$ (smoothly) when we switch off gravity. 
	The coefficients $A_\pm^{(i)}$ and $B_\pm^{(i)}$ are polynomial functions of $N_\text{f}$ and $f_2^2$, but their explicit forms are not relevant for our discussion.
	$\Delta^{(i)}$ is also a polynomial function of $N_\text{f}$ and $f_2^2$. Focusing on $N_\text{f} = 2$, we can write
	\begin{subequations}\label{eq:Deltas}
		\begin{equation}
			\begin{aligned}
				\Delta^{(1)} = \Delta^{(3)} & =  1342177280 \pi ^8 - 104857600 \pi ^6 f_2^2  \\
				& - 6471680 \pi ^4 f_2^4 + 634624 \pi ^2 f_2^6 \\
				& +  22445 f_2^8 \,,	 
			\end{aligned}
		\end{equation}
		\begin{equation}
			\begin{aligned}
				\Delta^{(2)} = \Delta^{(4)} &= 
				33554432000 \pi^8 - 2621440000 \pi^6 f_2^2 \\
				&- 201113600\pi^4 f_2^4 + 12572416\pi^2 f_2^6 \\
				&+ 561125f_2^8 \,.
			\end{aligned}
		\end{equation}
	\end{subequations}
	We can explicitly check that $\Delta^{(i)} > 0$ for positive values of $f_2^2$ (we are not interested in $f_2^2< 0$, as we cannot connect this region with asymptotically free trajectories). We explicitly checked that the same holds for $N_\text{f}>2$. 
	Therefore, we can conclude that all the fixed points of $\beta_{\lambda_\pm}$ remain real, even for large values of $f_2^2$, which allows us to define a chiral phase even at large values of $f_2^2$. We will discuss the phase structure in more detail once we lift the TT-approximation.   
	The picture we are developing here is qualitatively different from our QCD-example, where $\lambda_{\sigma,*}^\text{(IR)}$ and $\lambda_{\sigma,*}^\text{(UV)}$ move into the complex plane for large values of $g^2$.
	
	In Fig. \ref{fig:lambdafp _TTapprox}, we plot $\lambda_{\pm,*}^{(i)}$ ($i \in \{ 1,2,3,4 \}$) as a function of $f_2^2$.
	Increasing the value of $f_2^2$ we can see two fixed-point collisions, one for each pair of fixed points. 
	In contrast with the QCD-example, in the present case, the fixed points do not run into the complex plane after the fixed-point collision.
	As we shall see later, the fixed points involved in the fixed-point collision shown in Fig. \ref{fig:lambdafp _TTapprox} exchange their stability properties.	
	%%%%%%%%%%%%%%%%%%%%%%%%%%%%%%%%%%%%%%%%%%
	%%%%%%%%%%%%%%%%%%%%%%%%%%%%%%%%%%%%%%%%%%		
	Moving away from the TT-approximation, we restore all modes of $h_{\mu\nu}$. In this case, the flow of the four-fermion couplings is described by the beta function in \eqref{eq:betas_four-fermion}, where we treat $f_2^2$ and $f_0^2$ as external parameters. 
	The fixed points of $\beta_{\lambda_\pm}$ exhibit the same structure as Eq. \eqref{eq:struct_zeros}, but with different expressions for $A_\pm^{(i)}$, $B_\pm^{(i)}$ and $\Delta^{(i)}$, which now also depend on $f_0^2$. 
	
	In Fig. \ref{fig:DeltaContour_Nf2}, we show contour plots of $\Delta^{(i)}$ ($i=1,2$) as a function of $f_2^2$ and $f_0^2$ for $N_\text{f} =2$. 
	We focus on $f_2^2 > 0$ and $f_0^2 < 0$ since this is the region in the space of gravitational couplings where we can take values (at finite $k$) that are connected to the asymptotically free fixed point.
	We note that $\Delta^{(i)} > 0$ for all values of $f_2^2$ and $f_0^2$ in this range (we also explored a larger region of the parameter space and the result remains the same).
	Thus, we can conclude that the fixed points of $\beta_{\lambda_\pm}$ remain real even at large (absolute) values of the gravitational coupling $f_2^2$ and $f_0^2$, strengthening the results obtained with the TT-approximation.
	
	In Fig. \ref{fig:lambda_fp}, we plot $\lambda_{\pm,*}^{(i)}$ ($i \in \{ 1,2,3,4 \}$) as a function of $f_2^2$, for $f_0^2 = 0$.
	We also tested other values within the region $f_0^2<0$ and the qualitative conclusions reported here remain the same.
	Similarly to the TT-approximation, the fixed-point structure obtained with all graviton modes also exhibits two fixed-point collisions. Each one of these collisions defines a line in the $(f_2^2,f_0^2)$-plane. These lines are represented in Fig.~\ref{fig:DeltaContour_Nf2}, and they divide the $(f_2^2,f_0^2)$-plane into three regions, which we call $\mathcal{R}_1$, $\mathcal{R}_2$ and $\mathcal{R}_3$ (c.f. caption of Fig.~\ref{fig:DeltaContour_Nf2}). As we will see, these regions lead to different properties of the phase diagram of four-fermion interactions.
	
	The results we obtain with the inclusion of all graviton modes also indicate that we can define a chiral phase even at large (absolute) values of the gravitational couplings $f_2^2$ and $f_0^2$. 
	To confirm this statement, we now look at the phase diagram of four-fermion interactions.
	The qualitative result is very little affected by variations of $f_0^2$, thus, we focus on $f_0^2 = -f_2^2$.
	In Fig. \ref{fig:phase_withgrav_Nf=2}, we illustrate the general behavior with the choices of $f_2^2=20$, $f_2^2=40$ and $f_2^2=60$. These choices belong to $\mathcal{R}_1$, $\mathcal{R}_2$ and $\mathcal{R}_3$, respectively.
	We note that the fixed points $\text{\textbf{FP}}_1$ and $\text{\textbf{FP}}_2$ ($\text{\textbf{FP}}_3$ and $\text{\textbf{FP}}_4$) interchange their stability properties when we move from $\mathcal{R}_1$ to $\mathcal{R}_2$ ($\mathcal{R}_2$ to $\mathcal{R}_3$).
	In Table \ref{tab:stability}, we summarize the stability properties in the various regions of the $(f_2^2,f_0^2)$-plane. 
	%%%%%%%%%%%%%%%%%%%%%%%%%%%%%%%%%%%%%%
	\begin{center}
		\begin{table}[t]
			\begin{tabular}{|c|l|}
				\hline\hline
				\hspace*{.2cm}\multirow{3}{*}{Region $\mathcal{R}_1$} \hspace*{.2cm} 
				& \hspace*{.3cm}$\textbf{\text{FP}}_1$: IR-attractive                                                            \\ \cline{2-2} 
				& \hspace*{.3cm}$\textbf{\text{FP}}_2$ and $\textbf{\text{FP}}_3$: Mixed stability \hspace*{.3cm} \\ \cline{2-2} 
				& \hspace*{.3cm}$\textbf{\text{FP}}_4$: UV-attractive                                                            \\ \hline\hline
				\multirow{3}{*}{Region $\mathcal{R}_2$} 
				& \hspace*{.3cm} $\textbf{\text{FP}}_2$: IR-attractive                                                            \\ \cline{2-2} 
				& \hspace*{.3cm} $\textbf{\text{FP}}_1$ and $\textbf{\text{FP}}_3$:  Mixed stability \hspace*{.3cm} \\ \cline{2-2} 
				& \hspace*{.3cm} $\textbf{\text{FP}}_4$: UV-attractive                                                            \\ \hline\hline
				\multirow{3}{*}{Region $\mathcal{R}_3$} 
				& \hspace*{.3cm} $\textbf{\text{FP}}_1$: IR-attractive                                                            \\ \cline{2-2} 
				& \hspace*{.3cm} $\textbf{\text{FP}}_1$ and $\textbf{\text{FP}}_4$:  Mixed stability \hspace*{.3cm} \\ \cline{2-2} 
				& \hspace*{.3cm} $\textbf{\text{FP}}_3$: UV-attractive                                                            \\ \hline\hline
			\end{tabular}
			\caption{\footnotesize{Stability properties on the various regions of the parameter space $(f_2^2,f_0^2)$ in the multi-flavor case ($N_\text{f}\geq2$).
			We use the term ``mixed stability'' to indicate that a fixed point has one UV- and one IR-attractive direction.}}
			\label{tab:stability}
		\end{table}
	\end{center}
	%%%%%%%%%%%%%%%%%%%%%%%%%%%%%%%%%%%%%%	
	If $(f_2^2,f_0^2) \in \mathcal{R}_1$, we define the chiral phase as the basin of attraction of the fixed point $\textbf{\text{FP}}_1$. Meanwhile, if $(f_2^2,f_0^2) \in \mathcal{R}_2$ or $(f_2^2,f_0^2) \in \mathcal{R}_3$, we define the chiral phase as the basin of attraction of $\textbf{\text{FP}}_2$. The regions $\mathcal{R}_1$ and $\mathcal{R}_3$ share some features. In both cases, all fixed points are connected to the chiral phase. In contrast, in region $\mathcal{R}_2$ we cannot connect the fixed point $\textbf{\text{FP}}_4$ with points inside the chiral phase.
	Nevertheless, in all cases one can define a chiral phase, adding another piece of evidence that gravity does not trigger spontaneous \chSB{} in our setting. 	
	
	To complete the analysis, in Fig. \ref{fig:lambdatraject}, we plot the renormalization group trajectories $\lambda_{+}(k)$ and $\lambda_{-}(k)$ starting from the free fixed point in the UV, for several choices of boundary conditions of the gravitational coupling at the Planck scale.
	One can see that, as long as $f_2^2$ and $f_0^2$ remain finite at the Planck scale, the four-fermion couplings do not run into divergences in the trans-Planckian regime.
	As we run towards the IR, we can encounter a divergence in the flow of $\lambda_{+}(k)$ and $\lambda_{-}(k)$, which is generated by the Landau pole in $f_2^2(k)$ and $f_0^2(k)$. 
	This divergence is spurious since the Landau pole in the gravitational sector is most likely an indication of the breakdown of the beta functions \eqref{eq:betas_quadgrav_2a} and \eqref{eq:betas_quadgrav_2b}.
	In order to properly access the regime below the Planck scale, we need an IR completion of quadratic gravity. In the QCD-analogous scenario, the IR completion is an effective field theory with the Einstein-Hilbert action being the zeroth order term. However, up to the present moment, it is not clear whether there is a dynamical mechanism describing the transition from quadratic gravity (in the trans-Planckian regime) to an effective field theory (at low energies).
	
	Here, we explore a toy-model for the IR-completion where we set the running of the gravitational couplings according to 
	\begin{equation}\label{eq:toygravity}
		f_i^2(k) = 
		f_{i,\text{Pl.}}^2 \,\theta\big(k_\text{Pl.} - k\big) + 
		f_{i,\text{UV}}^2(k)\,\theta\big(k - k_\text{Pl.} \big) \,,
	\end{equation}
	with $i=0,2$. We use $f_{i,\text{UV}}^2(k)$ to parameterize the flow in the trans-Planckian regime (determined by \eqref{eq:betas_quadgrav_2a} and \eqref{eq:betas_quadgrav_2b}), and we define $f_{i,\text{Pl.}}^2 = f_{i,\text{UV}}^2(k_\text{Pl.})$. In this toy-model, the gravitational couplings are constant below the Planck scale.
	This toy-model is meant to illustrate that the flow of the four-fermion couplings does not run into divergences in the IR, therefore avoiding \chSB{}. 
	This situation, although simple, shows a significant difference with QCD, since a very similar toy-model applied to the single-channel approximation in QCD exhibits features related to \chSB{}. 
	In Fig. \ref{fig:lambdaGravQCD}, the left-panel shows the flow of four-fermion couplings under the influence of \eqref{eq:toygravity}, while the right-panel shows a the running of $\lambda_\sigma$ in our toy-model for the QCD case.
	
	In conclusion, the collection of results presented in this section shows evidence that in the QCD-analogous scenario for quadratic gravity, the non-perturbative sector does not lead to \chSB{} induced by quantum gravity effects.
	Therefore, the scenario proposed by Holdom and Ren \cite{Holdom:2015kbf,Holdom:2016xfn,Holdom:2019ouz} seems to be compatible with the existence of light fermions in our Universe.	
	
	%%%%%%%%%%%%%%%%%%%%%%%%%%%%%%%%%%%%%%
	%%%%%%%%%%%%%%%%%%%%%%%%%%%%%%%%%%%%%%
	%%%%%%%%%%%%%%%%%%%%%%%%%%%%%%%%%%%%%%
	\subsection{\chSB{} in Quadratic Gravity: $N_\text{f} =1$ \label{sec:chSBSingleFlavor}}
	\begin{figure}[t]
		\hspace*{-0.8cm}
		\includegraphics[width=.8\linewidth]{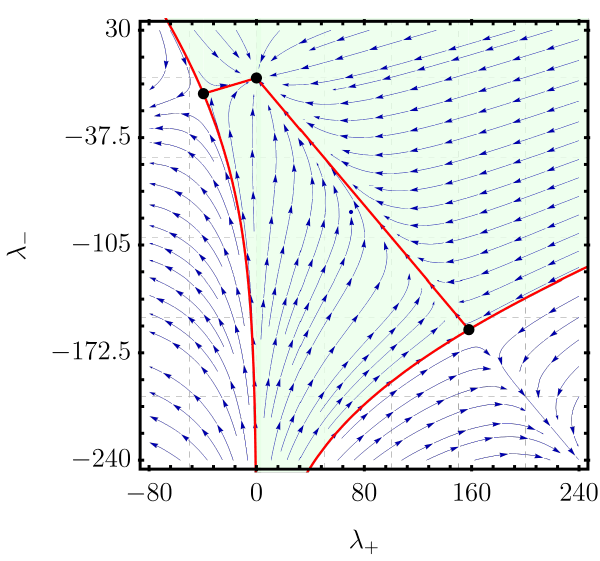}
		\caption{
			Phase diagram of four-fermion interaction in the single-flavor case $N_\text{f}=1$ and in the absence of gravitational contributions. 
			The green region indicates the chiral phase. Any UV-initial condition inside this region leads to trajectories that are attracted to the IR fixed point $\text{FP}_1^{(0)}$. 
			The red lines correspond to separatrix between the different sub-regions of the phase diagram.}
		\label{fig:phase_nograv_Nf=1}
	\end{figure}
	\begin{figure*}[t]
		\hspace*{-0.8cm}
		\includegraphics[width=.9\linewidth]{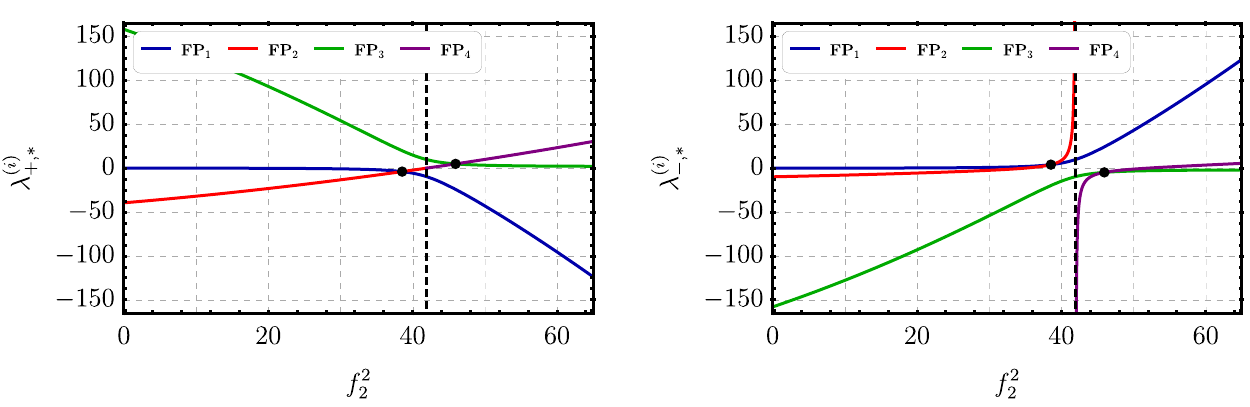}
		\caption{
			We show the fixed-point values $\lambda_{+,*}$ and $\lambda_{-,*}$ as a function of $f_2^2$ in the single-flavor case ($N_\text{f}=1$).
			In this plot, we use $f_0^2 = 0$ as a representative case.
			The black dots indicate fixed-point collisions, and the dashed lines mark the value of $f_2^2$ for which $\lambda_{-,*}$ tends to infinity. 
		}
		\label{fig:lambda_fp_Nf1}
	\end{figure*}
	\begin{figure*}[t]
		\includegraphics[width=\linewidth]{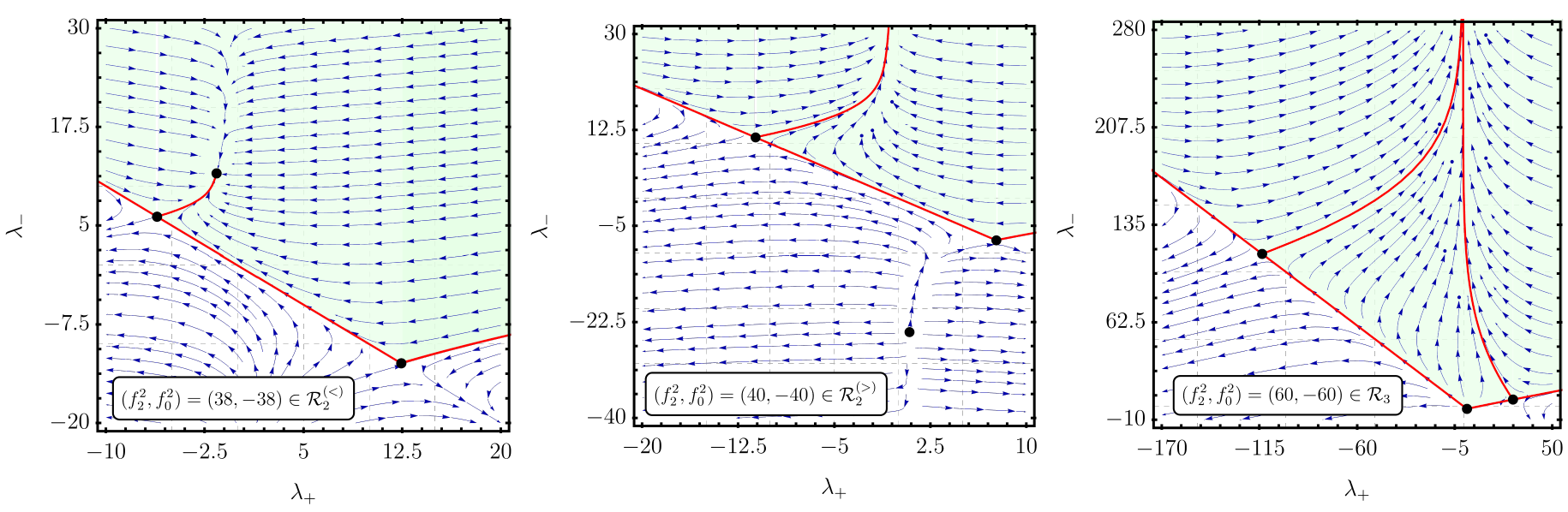}
		\caption{
			Phase diagram of four-fermion interaction in the single-flavor case ($N_\text{f}=1$) with the inclusion of gravitational contributions. 
			The green region indicates the chiral phase.
			In the second and third plots, we define the chiral phase as the basin of attraction of an IR fixed point lying at $\lambda_{-} \to + \infty$.}
		\label{fig:phase_withgrav_Nf=1}
	\end{figure*}
	\begin{figure*}[t]
		\hspace*{-0.8cm}
		\includegraphics[width=.85\linewidth]{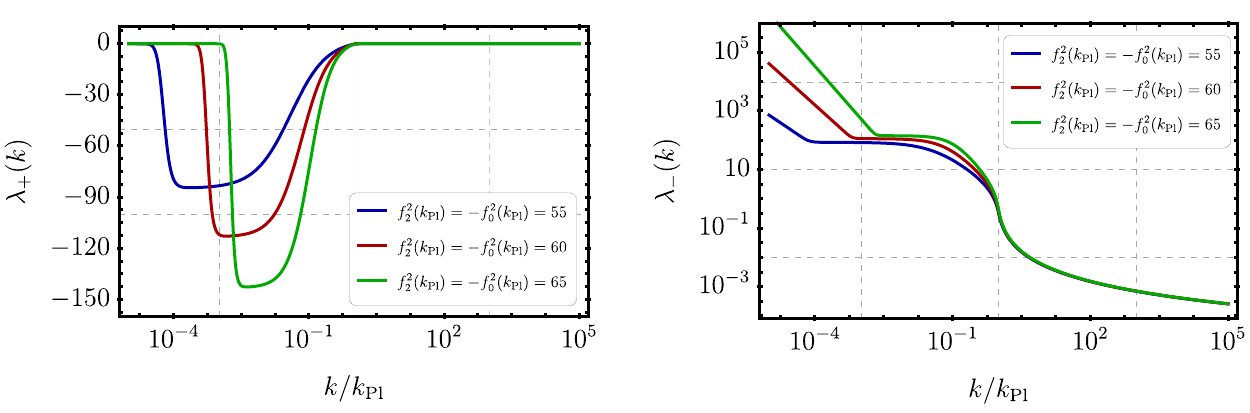}
		\caption{
			Renormalization group trajectories of $\lambda_\pm(k)$ in the single-flavor case ($N_\text{f}=1$).
			We consider Planck scale boundary conditions such that the gravitational couplings lie on region $\mathcal{R}_3$. With this choice, we see that four-fermion coupling $\lambda_-$ grows towards the IR. This growth reflects the fact that these trajectories are attracted to an IR fixed point with at infinite (in the $\lambda_{-}$-direction). 
		}
		\label{fig:lambdaGravNf1}
	\end{figure*}

	In this subsection, we look into the special case $N_\text{f} = 1$. As we shall see, this case exhibits a few differences in comparison with $N_\text{f} \geq 2$ and, therefore, deserves a separate analysis.
	
	The first difference between $N_\text{f}=1$ and $N_\text{f}\geq2$ happens already in the pure-fermion system, \textit{i.e.}, in the case where we switch off gravity.
	In this case, the beta functions $\beta_{\lambda_\pm}$ exhibit only three fixed points \cite{Braun:2011pp}, namely
	\begin{align}
		&\textbf{\text{FP}}_1^{(0)} : \quad (\lambda_{+,*},\lambda_{-,*}) = \left( 0 \, , \, 0  \right) \,, \\
		&\textbf{\text{FP}}_2^{(0)} : \quad(\lambda_{+,*},\lambda_{-,*}) = \left(- 4 \pi^2 \, , \, - \pi^2 \right)  \,, \\
		&\textbf{\text{FP}}_3^{(0)} : \quad(\lambda_{+,*},\lambda_{-,*}) = \left( 16\pi^2 \, , \, -16\pi^2 \right) \,.
	\end{align}
	As we can see in \eqref{eq:fixpt_PMint}, the would-be fixed point $\textbf{\text{FP}}_4^{(0)}$ runs to infinity when we take the limit $N_\text{f} \to 1$.
	The stability properties of these fixed points are the same as in the previous section.
	In Fig. \ref{fig:phase_nograv_Nf=1}, we plot the phase diagram of four-fermion couplings in the case $N_\text{f}=1$.
	As we did in the previous section, one can define the chiral region as the basin of attraction of  $\textbf{\text{FP}}_1^{(0)}$.
	The main difference in comparison with $N_\text{f} \geq 2$ (c.f.,~Fig. \ref{fig:phase_nograv_Nf=2}) is that, in the single flavor case, the chiral phase extends down to $\lambda_- \to -\infty$.
	%%%%%%%%%%%%%%%%%%%%%%%%%%%%%%%%%%%%%%
	\begin{center}
		\begin{table}[t]
			\begin{tabular}{|c|l|}
				\hline\hline
				\hspace*{.3cm}\multirow{2}{*}{Region $\mathcal{R}_1$} \hspace*{.3cm} 
				& \hspace*{.3cm}$\textbf{\text{FP}}_1$: IR-attractive                                                            \\ \cline{2-2} 
				& \hspace*{.3cm}$\textbf{\text{FP}}_2$ and $\textbf{\text{FP}}_3$: Mixed stability    \\ \hline\hline
				\multirow{2}{*}{Region $\mathcal{R}_2^{(<)}$} 
				& \hspace*{.3cm} $\textbf{\text{FP}}_2$: IR-attractive                                                            \\ \cline{2-2} 
				& \hspace*{.3cm} $\textbf{\text{FP}}_1$ and $\textbf{\text{FP}}_3$: Mixed stability     \\ \hline\hline
				\multirow{2}{*}{Region $\mathcal{R}_2^{(>)}$} 
				& \hspace*{.3cm} $\textbf{\text{FP}}_1$ and $\textbf{\text{FP}}_3$: Mixed stability \hspace*{.3cm} \\ \cline{2-2} 
				& \hspace*{.3cm} $\textbf{\text{FP}}_4$: UV-attractive                                                            \\ \hline\hline
				\multirow{2}{*}{Region $\mathcal{R}_3$} 
				& \hspace*{.3cm} $\textbf{\text{FP}}_1$ and $\textbf{\text{FP}}_4$: Mixed stability \hspace*{.3cm} \\ \cline{2-2} 
				& \hspace*{.3cm} $\textbf{\text{FP}}_3$: UV-attractive                                                            \\ \hline\hline
			\end{tabular}
			\caption{\footnotesize{Stability properties on the various regions of the parameter space $(f_2^2,f_0^2)$ in the single-flavor case ($N_\text{f}=1$).
			We use the term ``mixed stability'' to indicate that a fixed point has one UV- and one IR-attractive direction.}}
			\label{tab:stabilitySingFlav}
		\end{table}
	\end{center}
	%%%%%%%%%%%%%%%%%%%%%%%%%%%%%%%%%%%%%%	

	Switching on gravity and treating the gravitational couplings as external parameters, we first look at the structure of fixed points of $\beta_{\lambda_\pm}$ as a function of $f_2^2$ and $f_0^2$.
	Following the same notation as in the previous section, we use $\textbf{\text{FP}}_i$ to indicate the fixed points that are connected to $\textbf{\text{FP}}_i^{(0)}$ when we switch off gravitational contributions.
	To simplify the discussion, let us focus on the case $f_0^2 = 0$ (we explicitly verified that our conclusions remain unchanged for other values of $f_0^2$).
	In Fig. \ref{fig:lambda_fp_Nf1}, we plot the fixed-point values of $\lambda_+$ and $\lambda_-$ as a function of $f_2^2$. At this point, one can see a significant difference in comparison with the case $N_\text{f} \geq 2$.
	Starting at $f_2^2 = 0$, the fixed point $\textbf{\text{FP}}_2$ diverges (with $\lambda_- \to + \infty$) for a finite value $f_2^2 = f_{2,\text{div}}^2$ ($\approx 41.94$, for $f_0^2 = 0$).
	At the particular value $f_2^2 = f_{2,\text{div}}^2$, the beta functions $\beta_{\lambda_\pm}$ have only two fixed-point solutions with finite couplings.
	Slightly above $f_2^2 = f_{2,\text{div}}^2$, one sees a third fixed-point solution emerging from $\lambda_- \to - \infty$. However, this fixed point has the same properties as $\textbf{\text{FP}}_4$ instead of $\textbf{\text{FP}}_2$. In particular, one can check that this fixed-point solution corresponds to the limit $N_\text{f} \to 1$ of the fixed-point solution $\textbf{\text{FP}}_4$ described in the previous section.
	
	Besides the divergence at $f_2^2 = f_{2,\text{div}}^2$, the fixed-point structure of the case $N_\text{f}=1$ also exhibits two fixed-point collisions. The first, at $f_2^2 < f_{2,\text{div}}^2$, involves a collision between $\textbf{\text{FP}}_1$ and $\textbf{\text{FP}}_2$. The second, at $f_2^2 > f_{2,\text{div}}^2$, involves a collision between $\textbf{\text{FP}}_3$ and $\textbf{\text{FP}}_4$. In both cases, the fixed points involved in the collision interchange their stability properties.
	In this case, we can split the parameter space of gravitational couplings into four regions: $\mathcal{R}_1$, $\mathcal{R}_2^{(<)}$, $\mathcal{R}_2^{(>)}$ and $\mathcal{R}_3$. The regions $\mathcal{R}_1$ and $\mathcal{R}_2$ are similar to the corresponding ones that we introduced in the previous section. $\mathcal{R}_2^{(<)}$ and $\mathcal{R}_2^{(>)}$ are sub-divisions of the region $\mathcal{R}_2$ introduced in the previous section. $\mathcal{R}_2^{(<)}$ is below $f_2^2 = f_{2,\text{div}}^2$, while $\mathcal{R}_2^{(>)}$ is above $f_2^2 = f_{2,\text{div}}^2$. In Table \ref{tab:stabilitySingFlav}, we summarize the stability properties of the various regions in the parameter space of gravitational couplings.
	
	Because of the divergence of $\textbf{\text{FP}}_2$, we can only define the chiral phase based on a finite fixed point within region $\mathcal{R}_1$.
	For the other regions of the gravitational parameter space, the chiral phase corresponds to the basin of attraction of an IR fixed point located at infinity.
	In Fig. \ref{fig:phase_withgrav_Nf=1}, we show examples of the phase diagram obtained with $N_\text{f}=1$, within regions $\mathcal{R}_2^{(<)}$, $\mathcal{R}_2^{(>)}$ and $\mathcal{R}_3$. We refrain from plotting the phase diagram corresponding to region $\mathcal{R}_1$ since it does not show any qualitative difference from the results obtained with $N_\text{f}\geq2$.
	
	Finally, in Fig. \ref{fig:lambdaGravNf1}, we look at the renormalization group trajectories obtained with $N_\text{f} = 1$. Again, let us consider our toy-model for IR-completion where the running of gravitational couplings is given by Eq. \eqref{eq:toygravity}.
	At this point, we see a significant difference in comparison to the $N_\text{f} \geq 2$ case discussed in the previous section: if the Planck-scale values of $f_2^2$ and $f_0^2$ are sufficiently large, such that they belong to $\mathcal{R}_2^{(>)}$ or $\mathcal{R}_{3}$, the trajectories $\lambda_+(k)$ and $\lambda_-(k)$ grow unbounded when we flow towards the infrared.
	Nevertheless, this is not necessarily an indication of \chSB{}, since the coupling $\lambda_-$ would tend to infinity only at $k \to 0$, not at finite $k$ as in QCD.
	
	In conclusion, although the results of this section indicate that single-flavor case ($N_\text{f}=1$) exhibits a different phase structure in comparison with $N_\text{f} \geq2$, we remain without any indication of \chSB{} at finite renormalization group scale.	
	
	%%%%%%%%%%%%%%%%%%%%%%%%%%%%%%%%%%%%%%
	%%%%%%%%%%%%%%%%%%%%%%%%%%%%%%%%%%%%%%
	%%%%%%%%%%%%%%%%%%%%%%%%%%%%%%%%%%%%%%
	\subsection{Quadratic Gravity vs. QCD}
	\begin{figure*}[t]
		\includegraphics[width=\linewidth]{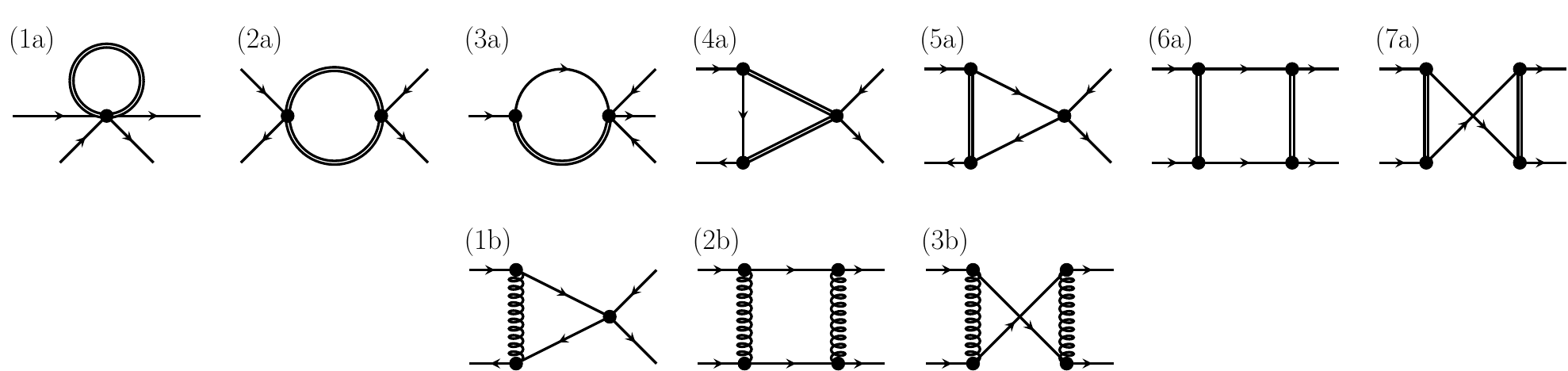}
		\caption{
			First row: Diagrams representing the gravitational contribution to the flow of four-fermion couplings in quadratic gravity.
			Second row: Diagrams representing the gluonic contributions to the flow of four-fermion couplings in QCD.
			These diagrams are FRG-diagrams, thus, the vertices and propagators are derived from $\Gamma_k$, not from the bare action. For simplicity, we omit the dependence on the regulator insertion $k \partial_k \textbf{R}_k$.
		}
		\label{fig:diagrams}
	\end{figure*}
	
	The results presented in this section go against the expectation that quadratic gravity, due to its attractive nature combined with a strongly correlated regime around the Planck scale, could trigger \chSB{}.
	In this subsection, we confront quadratic gravity with QCD to identify the aspects that differentiate both cases.
	The arguments presented here are based in \cite{Eichhorn:2011pc}, where a potential mechanism of gravitational induced \chSB{} was investigated in asymptotically safe quantum gravity.
	
	First, we note several differences between quadratic gravity and QCD at the level of the diagrams contributing to the flow of four-fermion interactions (c.f., Fig.~\ref{fig:diagrams}). 
	In the case of QCD, the gluon contributions to the flow of four-fermion interaction come from diagrams constructed in terms of the four-fermion vertex and the gluon-fermion-fermion vertex. 
	In quadratic gravity, there are infinitely many vertices involving fermions and metric fluctuations, allowing additional one-loop diagrams.
	
	Generically, we can trace the mechanism of \chSB{} in QCD back to the dominance of the $\lambda_i$-independent terms in $\beta_{\lambda_i}$ (diagrams 2b and 3b in Fig.~\ref{fig:diagrams}) in comparison with terms linearly proportional to $\lambda_i$ (diagram 1b in Fig.~\ref{fig:diagrams}).
	In the gravitational case, the situation is the opposite. Due to non-trivial cancellations between the box and crossed-box diagrams, the $\lambda_i$-independent terms in $\beta_{\lambda_i}$ (diagrams 2a, 4a, 6a and 7a in Fig.~\ref{fig:diagrams}) are over-weighted by the terms proportional to $\lambda_i$ (i.e., diagrams 1a, 3a, 5a in Fig.~\ref{fig:diagrams}, plus the anomalous dimension $\eta_\psi$), thus avoiding \chSB{} in quadratic gravity.
	
	In physical terms, the $\lambda_i$-independent terms in the flow of four-fermion interaction are related to the attractive nature of gravity. In contrast, the linear terms in $\lambda_\pm$ correspond to the anomalous scaling of four-fermion interactions.
	Our findings indicate that metric fluctuations in quadratic gravity contributes to the anomalous scaling of four-fermion couplings in such a way that it counteracts the attractive nature of gravity.
	Thus, the physical picture is similar to the results obtained in the context of asymptotically safe quantum gravity \cite{Eichhorn:2011pc}. 
	%%%%%%%%%%%%%%%%%%%%%%%%%%%%%%%%%%%%%%%%%%%%%%%%%%%%%%%%%%%%%%%%%%%%%%%%%%%%%%%%%%%%%%%%%%%	
	%%%%%%%%%%%%%%%%%%%%%%%%%%%%%%%%%%%%%%%%%%%%%%%%%%%%%%%%%%%%%%%%%%%%%%%%%%%%%%%%%%%%%%%%%%%	
	%%%%%%%%%%%%%%%%%%%%%%%%%%%%%%%%%%%%%%%%%%%%%%%%%%%%%%%%%%%%%%%%%%%%%%%%%%%%%%%%%%%%%%%%%%%	
	\section{Final remarks \label{sec:Conclusions}}
	
	In this paper, we explored the quantum gravity scenario conjectured by Holdom and Ren based on analogies between quadratic gravity and QCD.
	The goal of this paper was to put forward a first investigation concerning the phenomenological viability of this scenario, in particular, by confronting the existence of light fermions with a possible mechanism of \chSB{} around the Planck scale.
	
	We performed a renormalization group analysis of four-fermion interactions in order to investigate if quantum gravity effects could trigger \chSB{}. The main results of this paper are:
	
	\vspace*{.15cm}
	\noindent In a system containing two or more fermions ($N_\text{f}\geq2$), we found no indication for \chSB{} in connection with the non-perturbative regime of quadratic gravity.
	The results were obtained based on: 
	i) the fixed-point structure and general properties of the phase-diagram of four-fermion interactions; ii) finite renormalization group trajectories obtained by matching quadratic gravity in the UV with a toy-model for IR-completion.
	In general, we obtain a phase-diagram of four-fermion interactions that shows significant differences with the QCD-counterpart.
	
	\vspace*{.15cm}
	\noindent In a system containing only one fermion ($N_\text{f} = 1$), the phase structure of the four-fermion couplings behaves differently from the $N_\text{f}\geq2$ case. In particular, for sufficiently large Planck scale values of the gravitational couplings, we found that the four-fermion coupling grows unbounded towards the IR. However, these results do not constitute an indication for \chSB{} at finite renormalization group scale $k$.
	
	\vspace*{.15cm}
	The results of this paper indicate that the scenario proposed by Holdom and Ren passes the consistency test on whether it can accommodate light fermions in its landscape. 
	Nevertheless, there is still a considerable amount of work to be done in order to clarify whether the analogy between quadratic gravity and QCD can be dynamically realized.
	As a next step, we plan to investigate non-perturbative correlation functions in quadratic gravity, in particular, aiming to investigate whether 2-point functions exhibit the desired properties discussed in Sec. \ref{sec:QuadGrav_Rev}. 
	
	One of the attractive features of the analogy between quadratic gravity and QCD is the possibility of importing methods used in the context of QCD to access the non-perturbative regime of quadratic gravity.
	For example, functional methods such as FRG \cite{Pawlowski:2005xe,Gies:2006wv,Dupuis:2020fhh} and Dyson-Schwinger \cite{Roberts:1994dr,Huber:2018ned} equations have been very successful in the calculation of correlation functions in QCD. Thus it is conceivable that the same tools can be applied to the setup discussed in this paper. We plan to report on this in future papers.
	
	The setup discussed in the paper also resembles some of the ideas proposed by Donoghue in the context of Einstein-Cartan formulation of gravity \cite{Donoghue:2016vck} (see also \cite{Alexander:2022acv}). 
	In \cite{Donoghue:2016vck}, the author investigates whether there might be a mechanism of confinement/condensation involving the spin connection. The setup proposed by Donoghue also begs the question of whether \chSB{} can be triggered due to a strongly correlated gravitational regime. We plan to look into this direction in the future.
	
	The results presented here are also aligned with investigations of \chSB{} in the asymptotic safety approach to quantum gravity. In Ref.~\cite{Eichhorn:2011pc,Meibohm:2016mkp}, it was shown that a non-perturbative fixed point regime in quantum gravity does not trigger \chSB{}. More recently, the interplay between \chSB{}, gauge fields, and gravity was used to impose lower bounds on the number of fermions compatible with certain types of UV-completion with asymptotically safe quantum gravity \cite{deBrito:2020dta}.
	In the future, it would be interesting to explore this interplay within the context of quadratic gravity in analogy with QCD.
	
	%%%%%%%%%%%%%%%%%%%%%%%%%%%%%%%%%%%%%%%%%%%%%%%%%%%%%%%%%%%%%%%%%%%%%%%%%%%%%%%%%%%%%%%%%%%%
	\begin{acknowledgments}
		GPB would like to thank Astrid Eichhorn, Aaron Held, Benjamin Knorr,  Ant\^{o}nio Pereira, Shouryya Ray, Marc Schiffer and Arthur Vieira for interesting discussions on chiral symmetry breaking and/or quadratic gravity. 
		The author also thanks Astrid Eichhorn, Ant\^{o}nio Pereira, Marc Schiffer and Arthur Vieira for their feedback on the manuscript.
		GPB is supported by the research grant 29405 from VILLUM fonden.
	\end{acknowledgments}
	
	\bibliographystyle{unsrt}
	\bibliography{Refs}
\end{document}